\documentclass[10pt]{amsart}
\usepackage[small]{caption}
\usepackage{amsmath,amsfonts,amscd,amssymb,epsf, epsfig}\usepackage{graphicx}

\theoremstyle{definition}

\theoremstyle{remark} 
\numberwithin{equation}{section}
\newcommand{\Z}{{\mathbb{Z}}}

\newcommand{\C}{\mathbb{C}}

\newcommand{\R}{\mathbb{R}}

\newcommand{\pa}{\partial}

\newcommand{\vep}{\varepsilon}

\begin{document}

\title{Three dimensional Casimir piston for massive scalar fields}
\author{S.C. Lim}\address{ Faculty of Engineering,
Multimedia University, Jalan Multimedia, Cyberjaya, 63100, Selangor
Darul Ehsan, Malaysia.} \email{sclim@mmu.edu.my}

\author{L.P. Teo}\address{Faculty of Information
Technology, Multimedia University, Jalan Multimedia, Cyberjaya,
63100, Selangor Darul Ehsan, Malaysia.}\email{lpteo@mmu.edu.my}

 \maketitle

\begin{abstract}
We consider Casimir force acting on a three dimensional rectangular
piston due to a massive scalar field subject to periodic, Dirichlet
and Neumann boundary conditions. Exponential cut-off method is used
to derive the Casimir energy in the interior region and the exterior
region separated by the piston. It is  shown that the divergent
term of the Casimir force acting on the piston due to the interior
region cancels with that due to the exterior region, thus render a
finite well-defined Casimir force acting on the piston. Explicit
expressions for the total Casimir force acting on the piston is
derived, which show that the Casimir force is always attractive for
all the different boundary conditions considered. As a function of
$a$ --- the distance from the piston to the opposite wall, it is
found that the magnitude of the Casimir force behaves like $1/a^4$
when $a\rightarrow 0^+$ and decays exponentially when $a\rightarrow
\infty$. Moreover, the magnitude of the Casimir force is always a
decreasing function of $a$. On the other hand, passing from massless
to massive, we find that the effect of the mass is insignificant
when $a$ is small, but the magnitude of the force is decreased for large $a$ in the massive case.

\end{abstract}

\section{Introduction}

  Casimir effect associated with piston geometry has   attracted considerable interest since its introduction by Cavalcanti \cite{16} a few years ago. The main attraction of Casimir piston is that such a geometric setup can resolve a serious divergent problem that plagues the Casimir calculations. In the conventional calculations of Casimir force inside a confined region such as a rectangular box, the nontrivial contribution of vacuum energy outside the box has been ignored, and the surface divergent terms which depends on the dimensions and geometry are discarded to obtain finite results under the pretext of some regularization schemes. It has been shown that such surface divergence cannot be removed by renormalization of physical parameters of the theory \cite{13, b}.
 Calvacanti \cite{16} considered a new geometric
configuration -- a two dimensional rectangular piston, and showed
that for a massless scalar field subject to Dirichlet boundary
conditions, the divergent part of the Casimir force acting on the
piston due to the interior and exterior regions cancel with each
other and the net result is a finite attractive Casimir force
without the surface divergence. This result has stimulated an interest in
studying the Casimir force on pistons with different geometric setups
and boundary conditions. In particular, Hertzberg et al \cite{17}
have studied the  Casimir effect for electromagnetic fields in three
dimensional rectangular pistons with perfectly conducting walls. It
was found that the Casimir force is always attractive. This work was
generalized in \cite{18} where finite  temperature effect was  taken
into account and pistons with general cross sections were
considered. Around the same time, Marachevsky also carried out a
similar investigation using a different approach \cite{19, 20, 21}.
In \cite{9}, Edery studied the case of a massless scalar field with
Dirichlet boundary conditions for three dimensional rectangular
pistons. He also found that the force on the piston is always
attractive, in contrast to the fact that the regularized Casimir
force for a three dimensional rectangular cavity for massless scalar
field with Dirichlet boundary conditions can be attractive or
repulsive depending on the relative size of the cavity. This work
was later generalized in \cite{7} to  massless scalar fields with Dirichlet and Neumann boundary conditions in any dimensions. In all
the above scenarios, the Casimir force was found to be attractive.
In \cite{25},  Barton found out that when weakly reflecting
dielectric materials are used, the Casimir force on the three
dimensional piston can become repulsive when the plate separation is
sufficiently large. On the other hand, the work of Zhai and Li
\cite{26} showed that in the case of mixed boundary conditions (one
plate with Dirichlet boundary conditions and one with Neumann
boundary conditions), the Casimir force on a rectangular piston in
one, two and three dimensions is always repulsive. More recently,
Casimir effect has been investigated for electromagnetic fields with
perfect magnetic conditions in rectangular pistons of arbitrary
dimensions \cite{23} and for massless scalar fields with Dirichlet
boundary conditions in pistons inside space-time with extra
compactified dimensions \cite{24}.

  To the best of our knowledge,  no work has   been carried out on the Casimir effect for
massive scalar fields in the piston setting. In this work, we
consider a massive scalar field in a three dimensional rectangular
piston subject to periodic boundary conditions, Dirichlet boundary
conditions as well as Neumann boundary conditions.  We use
exponential cut-off method to compute the cut-off dependent Casimir
energy inside and outside the piston. We show that the sum of the
Casimir energies has a divergent part that does not depend on the
position of the piston, and therefore, the force acting on the
piston is finite, without any regularization. Explicit formulas for
the Casimir force are derived, which can be written as an infinite
convergent sums of Bessel functions. From the formulas, it is easy to
deduce that the Casimir force is always attractive under all the
different boundary conditions. We also proved a stronger result: as
a function of $a$ --- the separation between the piston and the
opposite wall, the magnitude of the Casimir force behaves like
$1/a^4$ when $a\rightarrow 0^+$ and it decays exponentially when
$a\rightarrow 0$. Moreover, the magnitude of the force is always
decreasing from infinity to 0. The mass effect to the Casimir force
is considered and it is found that the mass effect is significant
only when $a$ is large. Some numerical simulations have been carried
out.

\section{Cancelation of divergence in three dimensional Casimir piston}
Consider   a $3$-dimensional
rectangular piston in the form of  a rectangular cavity
$[0, L_1]\times [0,L_2]\times [0,L_3]$ separated by a  plane $x_1=a$
(the piston) into two regions: the interior region $[0, a]\times
[0,L_2]\times  [0,L_3]$ and the exterior region $[a, L_1]\times
[0,L_2]\times  [0,L_3]$ (see Figure 1). At the end we are going to
let $L_1\rightarrow$ so that the exterior region becomes an open
region.  We want to compute the Casimir force acting on the piston
due to  a massive ($m>0$)
 scalar field $\phi(\mathbf{x}, t)$ satisfying the Klein--Gordon
equation in Minkowski space-time:
\begin{align}\label{eq6_19_1}
\left(\square +m^2\right)\phi(\mathbf{x}, t)=0, \hspace{1cm}
(\mathbf{x}, t)\in \R^{3+1},
\end{align}where $$\square =\frac{\pa^2}{\pa t^2} -\sum_{i=1}^3
\frac{\pa^2}{\pa x_i^2}.$$   The
periodic boundary condition (bc), Dirichlet bc and Neumann bc will
be considered.
\begin{figure} \epsfxsize=.6\linewidth
\epsffile{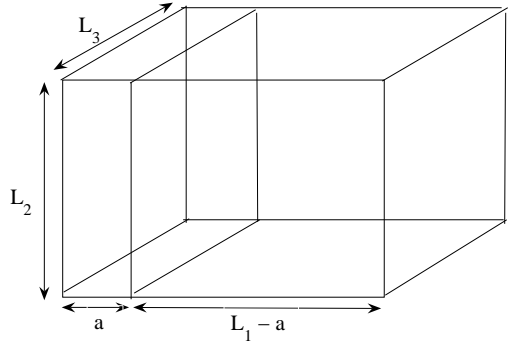}\caption{The three dimensional rectangular
pistons}\end{figure}

The Casimir energy of the piston system is the sum of the Casimir
energy of the internal region $E_{\text{cas}}(a, L_2, L_3; m)$ and
the Casimir energy of the external region $E_{\text{cas}}(L_1-a,
L_2, L_3; m)$. The Casimir force acting on the piston is then
obtained by
\begin{align}\label{eq6_20_4}
F_{\text{Cas}}=\lim_{L_1\rightarrow \infty}-\frac{\pa}{\pa
a}\Bigl(E_{\text{Cas}}(a, L_2, L_3; m) + E_{\text{Cas}}(L_1-a, L_2,
L_3;m)\Bigr).
\end{align}
In the periodic bc case, the eigenmodes of the $(d+1)$ --
dimensional field $\phi(\mathbf{x}, t)$ in a $d$-dimensional
rectangular cavity $[0, L_1]\times \ldots \times [0, L_d]$
satisfying \eqref{eq6_19_1} (with $3$ replaced by $d$) are given by
\begin{align*}
\omega_{\mathbf{k}}^P= \sqrt{ \sum_{i=1}^d\left[\frac{
2\pi}{L_i}\right]^2 +m^2}, \hspace{1cm}\mathbf{k}\in \Z^d;
\end{align*}and the Casimir energy is defined as the divergent sum
\begin{align}\label{eq6_19_2}
E_{\text{Cas}}^{P}(L_1, \ldots, L_d;
m)=\frac{1}{2}\sum_{\mathbf{k}\in \Z^d} \omega_{\mathbf{k}}^P.
\end{align}
For the Dirichlet bc and Neumann bc cases, the eigenmodes are
respectively
\begin{align*}
\omega_{\mathbf{k}}^D =\sqrt{ \sum_{i=1}^d\left[\frac{
\pi}{L_i}\right]^2 +m^2}, \hspace{1cm}\mathbf{k}\in \mathbb{N}^d;
\end{align*}and
\begin{align*}
\omega_{\mathbf{k}}^N =\sqrt{ \sum_{i=1}^d\left[\frac{
\pi}{L_i}\right]^2 +m^2}, \hspace{1cm}\mathbf{k}\in
(\mathbb{N}\cup\{0\})^d.
\end{align*}Therefore when $d=3$ \cite{5},
\begin{align}\label{eq6_20_2}
E_{\text{Cas}}^{D/N}(L_1, L_2, L_3; m)=&\frac{1}{8}\Biggl\{
E_{\text{Cas}}^{P}(2L_1, 2L_2, 2L_3; m)\\&\mp\Bigl[
E_{\text{Cas}}^{P}(2L_1, 2L_2; m)+ E_{\text{Cas}}^{P}(2L_1, 2L_3;
m)+ E_{\text{Cas}}^{P}(2L_2, 2L_3; m)\Bigr]\nonumber\\
&+E_{\text{Cas}}^{P}(2L_1;m)+E_{\text{Cas}}^{P}(2L_2;m)+E_{\text{Cas}}^{P}(2L_3;m)\mp
\frac{1}{2}m\Biggr\}\nonumber.
\end{align}Different
regularization schemes have been employed to define a regularized
Casimir energy from \eqref{eq6_19_2}. We adopt here the exponential
cut-off method which allows us to retain the divergent terms. Define the
$\lambda$-dependent Casimir energy by
\begin{align}\label{eq6_19_3}
E_{\text{Cas}}^{P}(\lambda; L_1, \ldots,
L_d;m)=\frac{1}{2}\sum_{\mathbf{k}\in \Z^d}
\omega_{\mathbf{k}}^Pe^{-\lambda \omega_{\mathbf{k}}^P}.
\end{align} In the Appendix A, we show that
\begin{align}\label{eq6_19_6}
E_{\text{Cas}}^P\left(\lambda; L_1, \ldots,  L_d;m\right)
=&E_{\text{Cas, div}}^P\left(\lambda; L_1,  \ldots,  L_d;m\right) +
E_{\text{Cas, reg}}^P(L_1,  \ldots,  L_d;m)+O(\lambda),
\end{align}where for $d=1, 2$ and $3$, the $\lambda\rightarrow 0^+$ divergent
term $E_{\text{Cas, div}}^P\left(\lambda; L_1,  \ldots, L_d;m\right)
$ and the regularized Casimir energy $E_{\text{Cas, reg}}^P(L_1,
\ldots,  L_d;m)$ are given respectively by
\begin{align*}
E_{\text{Cas, div}}^P(\lambda; L_1; m) =&
\frac{L_1}{2\pi}\lambda^{-2}-\frac{  m^2 L_1}{4\pi}\log\lambda;
\end{align*}
\begin{align}\label{eq6_20_6}
E_{\text{Cas, reg}}^P(\lambda; L_1; m) =& -\frac{ m^2
L_1}{4\pi}\left(\log m -\log 2-\frac{1}{4}+\gamma\right)-\frac{
m}{\pi}\sum_{k=1}^{\infty} k^{-1}K_1(mkL_1);
\end{align}
\begin{align*}
E_{\text{Cas, div}}^P(\lambda; L_1,
L_2;m)=&\frac{L_1L_2}{2\pi}\lambda^{-3};
\end{align*}
\begin{align}\label{eq6_20_7}
E_{\text{Cas, reg}}^P(\lambda; L_1,
L_2;m)=&-\frac{L_1L_2}{12\pi}m^3-\frac{L_1L_2}{\sqrt{8\pi^3}} m^{3/2}\\
&\times\sum_{\mathbf{k}\in \Z^2\setminus\{
\mathbf{0}\}}\left(\sum_{i=1}^2
\left[k_iL_i\right]^2\right)^{-\frac{3}{4}} K_{3/2} \left( m
\sqrt{\sum_{i=1}^2 \left[k_iL_i\right]^2}\right);\nonumber
\end{align}
\begin{align*}%\label{eq6_19_8}
E_{\text{Cas, div}}^P\left(\lambda; L_1, L_2, L_3;m\right)=\frac{3
L_1L_2L_3}{2\pi^2}\lambda^{-4}-\frac{L_1L_2L_3}{8\pi^2}m^2\lambda^{-2}+\frac{L_1L_2L_3}{32\pi^2}m^4
\log \lambda ;\end{align*}
\begin{align}\label{eq6_20_8}
E_{\text{Cas, reg}}^P(L_1, L_2, L_3;m)=&\frac{L_1L_2L_3}{32\pi^2}
m^4 \left( -\frac{1}{2}+\gamma+\log m-\log 2
\right)-\frac{L_1L_2L_3}{4\pi^2} m^2\\& \times\sum_{\mathbf{k}\in
\Z^3\setminus\{ \mathbf{0}\}}\left(\sum_{i=1}^3
\left[k_iL_i\right]^2\right)^{-1} K_{2} \left( m \sqrt{\sum_{i=1}^3
\left[k_iL_i\right]^2}\right).\nonumber
\end{align} Here $\gamma$ is the Euler constant and $K_{\nu}(z)$ is the modified Bessel function.
From these formulas and \eqref{eq6_20_2}, we can compute the cut-off
dependent Casimir energies for massive scalar fields inside three
dimensional rectangular cavities under various boundary conditions.
Conventionally, a finite Casimir energy is defined by taking
$\lambda\rightarrow 0^+$ limit in the regular term of the Casimir
energy. We would like to comment that the result will agree with the
result derived by zeta regularization method (see Appendix B) if one
defines the normalization constant $\mu$ appearing in the zeta
regularization method as $e^{-\gamma}$.

Now we analyze the divergent term of  the Casimir energy. In the
periodic bc case, we observe that the divergent term depends
linearly on the volume $V=L_1L_2L_3$ of the rectangular cavity.
However, for the cases of Dirichlet bc and Neumann bc,
\eqref{eq6_20_2} shows that the divergent term can be written in
the form
\begin{align}\label{eq6_20_3}
&E_{\text{Cas, div}}^{D/N}(\lambda; L_1, L_2, L_3;  m) \\=&
\mathfrak{D}_V(\lambda; m) L_1L_2L_3 \pm \mathfrak{D}_S(\lambda;
m)(L_1L_2+L_2L_3+L_1L_3) +\mathfrak{D}_L(\lambda;
m)(L_1+L_2+L_3).\nonumber
\end{align}In other words, besides the bulk divergence $\mathfrak{D}_V(\lambda; m)
L_1L_2L_3$ which only depends on the volume, there are also surface
divergence $\mathfrak{D}_S(\lambda; m)(L_1L_2+L_2L_3+L_1L_3) $ and
divergence due to corners $\mathfrak{D}_L(\lambda; m)(L_1+L_2+L_3)$.

For  the piston geometry, it is easy to see from
\eqref{eq6_20_3} that the divergent term for the sum of the Casimir
energy of the interior region and the exterior region given by
$E_{\text{Cas}}^{P/D/N}(\lambda; a, L_2, L_3; m)+
E_{\text{Cas}}^{P/D/N}(\lambda; L_1-a, L_2, L_3; m)$ depends only on
$L_1, L_2, L_3$, but not on the position of the piston $x_1=a$. In
other words, the divergent parts of the Casimir energies in the
interior and exterior regions contribute Casimir force of same
magnitude but opposite signs to the piston, and thus cancel with
each other. \emph{This renders a finite quantity for the Casimir
force acting on the piston} given by \eqref{eq6_20_4}, which can be
rewritten as
\begin{align}\label{eq6_20_4_1}
F_{\text{Cas}}^{P/D/N}=\lim_{L_1\rightarrow \infty}-\frac{\pa}{\pa
a}\Bigl(E_{\text{Cas, reg}}^{P/D/N}(\lambda=0; a, L_2, L_3; m) +
E_{\text{Cas, reg}}^{P/D/N}(\lambda=0;L_1-a, L_2, L_3;m)\Bigr).
\end{align}

\section{Analysis of the Casimir force}

From \eqref{eq6_20_6}, \eqref{eq6_20_7} and \eqref{eq6_20_8}, one
observes that the regularized Casimir energy $E_{\text{Cas,
reg}}^{P}(\lambda=0; L_1,\ldots, L_d;m)$, $d=1, 2, 3$ inside a
rectangular cavity can be written as a term $\mathfrak{B}_{d}(L_2,
\ldots, L_d; m)L_1$ which depends linearly on $L_1$, plus a Bessel
series
\begin{align*} R_d(L_1, \ldots,L_d; m)=&-
\frac{\prod_{i=1}^dL_i}{(2\pi)^{\frac{d+1}{2}}}m^{\frac{d+1}{2}}\sum_{\substack{k_1\in
\Z\setminus\{0\}\\ (k_2, \ldots, k_d) \in
\Z^{d-1}}}\left(\sum_{i=1}^d
\left[k_iL_i\right]^2\right)^{-\frac{d+1}{4}} \\
&\times K_{\frac{d+1}{2}} \left( m \sqrt{\sum_{i=1}^d
\left[k_iL_i\right]^2}\right).
\end{align*}Notice that the summation over $\mathbf{k}\in \Z^d\setminus\{0\}$ in
\eqref{eq6_20_6}, \eqref{eq6_20_7} and \eqref{eq6_20_8} is
decomposed into summation  over $k_1=0, (k_2, \ldots, k_d) \in
\Z^{d-1}\setminus\{0\}$ and summation over $k_1\neq 0, (k_2, \ldots,
k_d)\in\Z^{d-1}$. The term that involves summation over $k_1=0,
(k_2, \ldots, k_d) \in \Z^{d-1}\setminus\{0\}$ depends on $L_1$
linearly and is therefore  combined into  the term
$\mathfrak{B}_{d}(L_2, \ldots, L_d; m)L_1$. As in the case of the
divergent terms, for periodic bc as well as Dirichlet bc and
Neumann bc, the sum of the terms $\mathfrak{B}_d(L_2,\ldots,
L_d;m)L_1$ of the interior region and exterior region does not
depend on the position of the piston and thus the corresponding
Casimir force cancels with each other. In other words, the Casimir
force acting on the three dimensional piston is given by
\begin{align}\label{eq6_27_2}
F_{\text{Cas}}^P(a; L_2, L_3; m) = -\lim_{L_1\rightarrow
\infty}\frac{\pa}{\pa a}\Bigl\{R_3(a, L_2 ,L_3; m)+R_3(L_1-a, L_2,
L_3; m)\Bigr\}
\end{align}in the case of periodic bc; whereas for Dirichlet and
Neumann bc, \eqref{eq6_20_2} shows that
\begin{align}\label{eq6_27_3}
&F_{\text{Cas}}^{D/N}(a; L_2, L_3;
m)\\=&-\frac{1}{8}\lim_{L_1\rightarrow \infty}\frac{\pa}{\pa
a}\Bigl\{R_3(2a, 2L_2 ,2L_3; m)+R_3(2L_1-2a, 2L_2, 2L_3;
m)\nonumber\\&\mp R_2(2a, 2L_2;m) \mp R_2(2L_1-2a; 2L_2; m) \mp
R_2(2a, 2L_3;m)\nonumber\\&\mp R_2(2L_1-2a, 2L_3; m) + R_1(2a;
m)+R_1(2L_1-2a; m)\Bigr\}\nonumber.
\end{align}Using the formula (\cite{8}, \#3.478, no. 4),
\begin{align}\label{eq6_23_5}
\int_0^{\infty} t^{\pm \nu-1} \exp\left( -\alpha t
-\frac{\beta}{t}\right) dt = 2\left(\frac{\beta}{\alpha}\right)^{\pm
\frac{\nu}{2}}K_{\nu}(2\sqrt{\alpha\beta}),
\end{align}
we have
\begin{align}\label{eq6_20_10}
R_d(L_1, \ldots,L_d; m) =-
\frac{\prod_{i=1}^dL_i}{2^{d+2}\pi^{\frac{d+1}{2}}}\int_0^{\infty}
t^{-\frac{d+3}{2}}\sum_{\substack{k_1\in \Z\setminus\{0\}\\ (k_2,
\ldots, k_d) \in \Z^{d-1}}}\exp\left( - tm^2
-\frac{1}{4t}\sum_{i=1}^d \left[k_iL_i\right]^2\right)dt.
\end{align}From here, we see that $R_d(L_1, \ldots, L_d; m)$ decays
exponentially as $L_1\rightarrow \infty$. Taking derivative with
respec to $a$ and letting $L_1\rightarrow \infty$, we have
\begin{align*}
\lim_{L_1\rightarrow \infty}-\frac{\pa}{\pa a}R_d(L_1-a, \ldots,L_d;
m)=0.
\end{align*}In other words, there is no contribution to the Casimir force
on the piston from the terms $R_d(L_1-a, L_2,\ldots,L_d; m)$ of the
exterior region. Therefore, the net contribution to the Casimir
force on the piston are due to the terms $R_d(a, L_2,\ldots,L_d; m)$
of the interior region. More precisely,
\begin{align}\label{eq6_23_2}
F_{\text{Cas}}^P(a; L_2, L_3; m)=-\frac{\pa}{\pa a}R_3(a, L_2,L_3;
m),
\end{align}and
\begin{align}\label{eq6_23_3}
F_{\text{Cas}}^{D/N}(a; L_2, L_3;
m)=&\frac{1}{8}\Biggl\{-\frac{\pa}{\pa a}R_3(2a, 2L_2,2L_3;
m)\pm\frac{\pa}{\pa a} R_2(2a; 2L_2;m)\\& \pm\frac{\pa}{\pa
a}R_2(2a;2L_3;m) -\frac{\pa}{\pa a} R_1(2a; m)\Biggr\}.\nonumber
\end{align}
Now applying the Jacobi inversion formula
\begin{align}\label{eq6_23_1}
\sum_{k=-\infty}^{\infty} e^{-tk^2} =
\sqrt{\frac{\pi}{t}}\sum_{k=-\infty}^{\infty}
e^{-\frac{\pi^2k^2}{t}}
\end{align}to the summation over $(k_2, \ldots, k_d)\in
\Z^{d-1}$ in \eqref{eq6_20_10}, we find that
\begin{align}\label{eq6_24_6}
&R_d(a, L_2,\ldots,L_d; m) \\=&- \frac{a}{8\pi}\int_0^{\infty}
t^{-2}\sum_{\substack{k_1\in \Z\setminus\{0\}\\ (k_2, \ldots, k_d)
\in \Z^{d-1}}}\exp\left( - tm^2 -4t\pi^2\sum_{i=2}^d
\left[\frac{k_i}{L_i}\right]^2-\frac{1}{4t}[ak_1]^2\right)dt\nonumber\\
=&-\frac{1}{\pi} \sum_{k_1=1}^{\infty}\sum_{(k_2, \ldots,
k_d)\in\Z^{d-1}}\frac{1}{k_1} \sqrt{m^2+\sum_{i=2}^d\left[\frac{2\pi
k_i}{L_i}\right]^2}K_1\left(ak_1
\sqrt{m^2+\sum_{i=2}^d\left[\frac{2\pi
k_i}{L_i}\right]^2}\right).\nonumber
\end{align}Eq. \eqref{eq6_23_2} then shows that in the periodic bc case,
\begin{align*}
&F_{\text{Cas}}^P(a; L_2, L_3; m)\\=&\frac{1}{\pi}
\sum_{k_1=1}^{\infty}\sum_{(k_2, k_3)\in\Z^{2}}\frac{1}{k_1}
\sqrt{m^2+\sum_{i=2}^3\left[\frac{2\pi
k_i}{L_i}\right]^2}\frac{\pa}{\pa a}K_1\left(ak_1
\sqrt{m^2+\sum_{i=2}^3\left[\frac{2\pi k_i}{L_i}\right]^2}\right).
\end{align*}
On the other hand, using the fact that if $f(k_2, k_3)$ is a totally
even function, i.e. $f( \pm k_2, \pm k_3)= f( k_2, k_3)$, then$$
\frac{1}{4}\left\{\sum_{(k_2, k_3)\in \Z^2}\mp\sum_{k_2\in\Z,
k_3=0}\mp \sum_{k_2=0, k_3\in\Z} +\sum_{k_2=0, k_3=0}\right\} f(k_2,
k_3) = \sum_{(k_2, k_3)\in \mathbb{N}^2/(\mathbb{N}\cup\{0\})^2}
f(k_2,k_3);
$$Eqs. \eqref{eq6_23_3} and \eqref{eq6_24_6} give the expression
\begin{align*}
&F_{\text{Cas}}^{D/N}(a; L_2, L_3; m)\\=&\frac{1}{2\pi}
\sum_{k_1=1}^{\infty}\sum_{(k_2,
k_3)\in\mathbb{N}^{2}/(\mathbb{N}\setminus\{0\})^2}\frac{1}{k_1}
\sqrt{m^2+\sum_{i=2}^3\left[\frac{\pi
k_i}{L_i}\right]^2}\frac{\pa}{\pa a}K_1\left(2ak_1
\sqrt{m^2+\sum_{i=2}^3\left[\frac{\pi k_i}{L_i}\right]^2}\right).
\end{align*}for the Casimir force in the Dirichlet bc and Neumann bc
cases. Now using the formula [\cite{8}, \#8.486, no. 12]
\begin{align}\label{eq6_25_3}\frac{d}{dz}K_{\nu}(z)= -\frac{\nu}{z} K_{\nu}(z)
-K_{\nu-1}(z),\end{align} we obtain finally the explicit expression
for the Casimir force in the periodic bc, Dirichlet bc and Neumann
bc cases:
\begin{align}\label{eq6_25_2}
&F_{\text{Cas}}^P(a; L_2, L_3;m) \\=&-\frac{1}{\pi a}
\sum_{k_1=1}^{\infty}\sum_{(k_2, k_3)\in\Z^{2}}\frac{1}{k_1}
\sqrt{m^2+\sum_{i=2}^3\left[\frac{2\pi
k_i}{L_i}\right]^2}K_1\left(ak_1
\sqrt{m^2+\sum_{i=2}^3\left[\frac{2\pi k_i}{L_i}\right]^2}\right)\nonumber\\
&-\frac{1}{\pi } \sum_{k_1=1}^{\infty}\sum_{(k_2,
k_3)\in\Z^{2}}\left(m^2+\sum_{i=2}^3\left[\frac{2\pi
k_i}{L_i}\right]^2\right) K_0\left(ak_1
\sqrt{m^2+\sum_{i=2}^3\left[\frac{2\pi
k_i}{L_i}\right]^2}\right);\nonumber
\end{align}
\begin{align}\label{eq6_24_5}
&F_{\text{Cas}}^{D/N}(a; L_2, L_3; m)\\
=&-\frac{1}{2\pi a}\sum_{k_1=1}^{\infty}\sum_{ (k_2, k_3) \in
\mathbb{N}^{2}/
(\mathbb{N}\cup\{0\})^2}\frac{1}{k_1}\sqrt{m^2+\sum_{i=2}^3\left[\frac{\pi
k_i}{L_i}\right]^2}K_1\left(2a
k_1\sqrt{m^2+\sum_{i=2}^3\left[\frac{\pi
k_i}{L_i}\right]^2}\right)\nonumber\\
&-\frac{1}{\pi} \sum_{k_1=1}^{\infty}\sum_{ (k_2, k_3) \in
\mathbb{N}^{2}/
(\mathbb{N}\cup\{0\})^2}\left(m^2+\sum_{i=2}^3\left[\frac{\pi
k_i}{L_i}\right]^2\right) K_0\left(2a
k_1\sqrt{m^2+\sum_{i=2}^3\left[\frac{\pi
k_i}{L_i}\right]^2}\right).\nonumber
\end{align}
These formulas are useful for studying the large--$m$ and large--$a$
behavior of the Casimir force. More precisely, since the modified
Bessel function $K_{\nu}(z)$ decays exponentially as
$z\rightarrow\infty$, with leading term
\begin{align*}
K_{\nu}(z) \sim \sqrt{\frac{\pi}{2z}}e^{-z},
\end{align*} Eqs. \eqref{eq6_25_2} and \eqref{eq6_24_5} show that for fixed $m, L_2$ and  $L_3$, in the   large
$a$ ($a\gg 1$)  limit, the Casimir force $F_{\text{Cas}}^{P/D/N}(a;
L_2, L_3; m)$ decays exponentially with leading terms being
\begin{align*}
&F_{\text{Cas}}^{P}(a;  L_2, L_3; m) \sim -m\sqrt{\frac{2m}{\pi
a}}e^{-am}, \\&F_{\text{Cas}}^{N}(a;  L_2, L_3; m) \sim
-\frac{m}{2}\sqrt{\frac{m}{\pi a}}e^{-2am},\\
&F_{\text{Cas}}^{D}(a;  L_2, L_3; m) \sim
-\frac{1}{2}\sqrt{\frac{1}{\pi
a}}\left(m^2+\frac{\pi^2}{L_2^2}+\frac{\pi^2}{L_3^2}\right)^{3/4}\exp\left(-2a\sqrt{m^2+\frac{\pi^2}{L_2^2}
+\frac{\pi^2}{L_3^2}}\right)
\end{align*}respectively. In other words, for fixed $L_2$ and $L_3$, when the
piston is very far away from the opposite wall, the Casimir force is
an exponentially weak attractive force. In the massless
 $m \rightarrow 0^+$ limit, the Casimir force $F_{\text{Cas}}^D(a; L_2,
 L_3;0)$ for Dirichlet bc still decays exponentially with respect to
 $a$.
 However,  the Casimir
force $F_{\text{Cas}}^P(a; L_2, L_3;0)$ and $F_{\text{Cas}}^N(a;
L_2, L_3;0)$ for periodic bc and Neumann bc becomes  polynomially
decay with order $1/a^2$.

\begin{figure}\centering
\epsfxsize=.49\linewidth
\epsffile{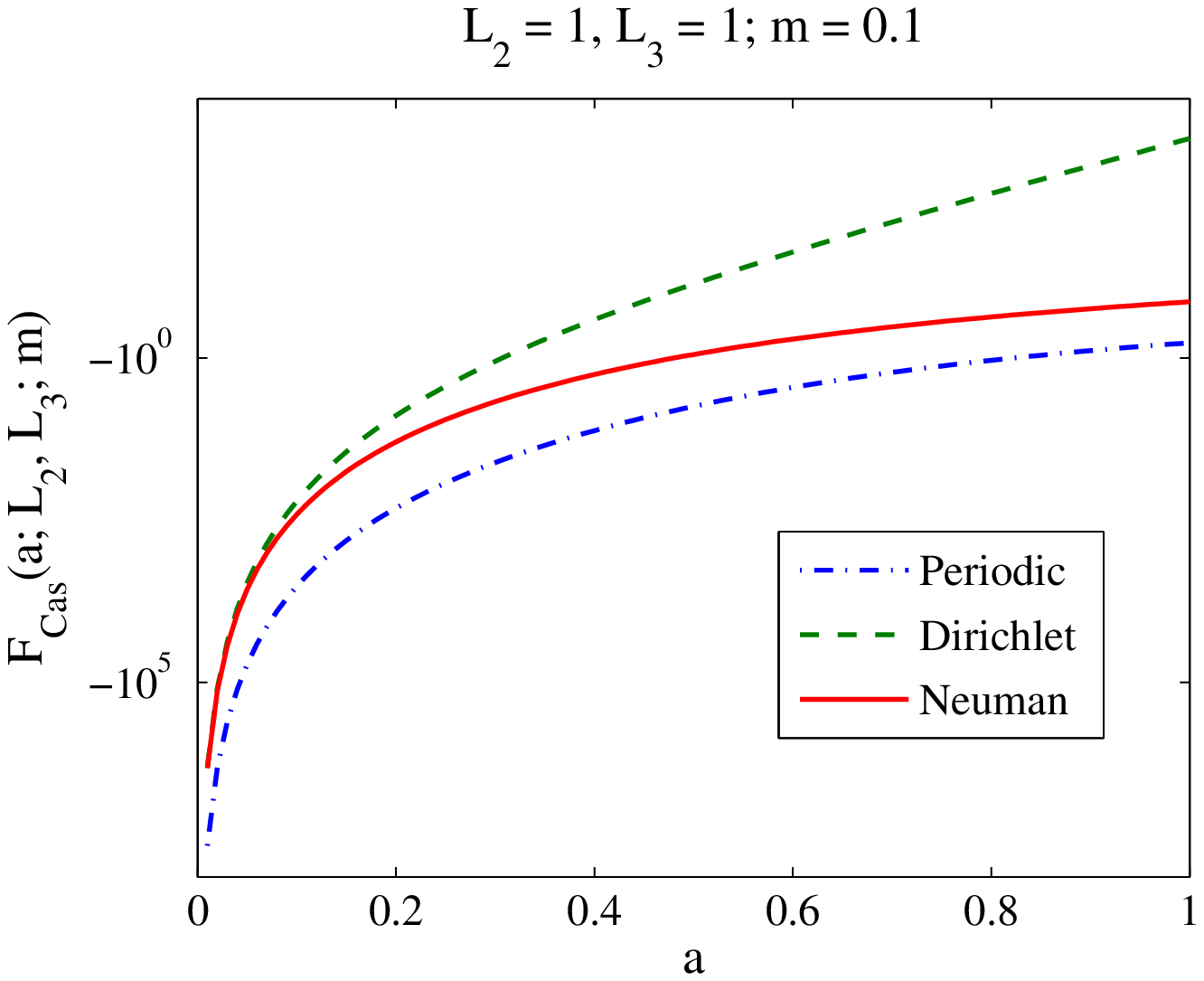}\epsfxsize=.49\linewidth
\epsffile{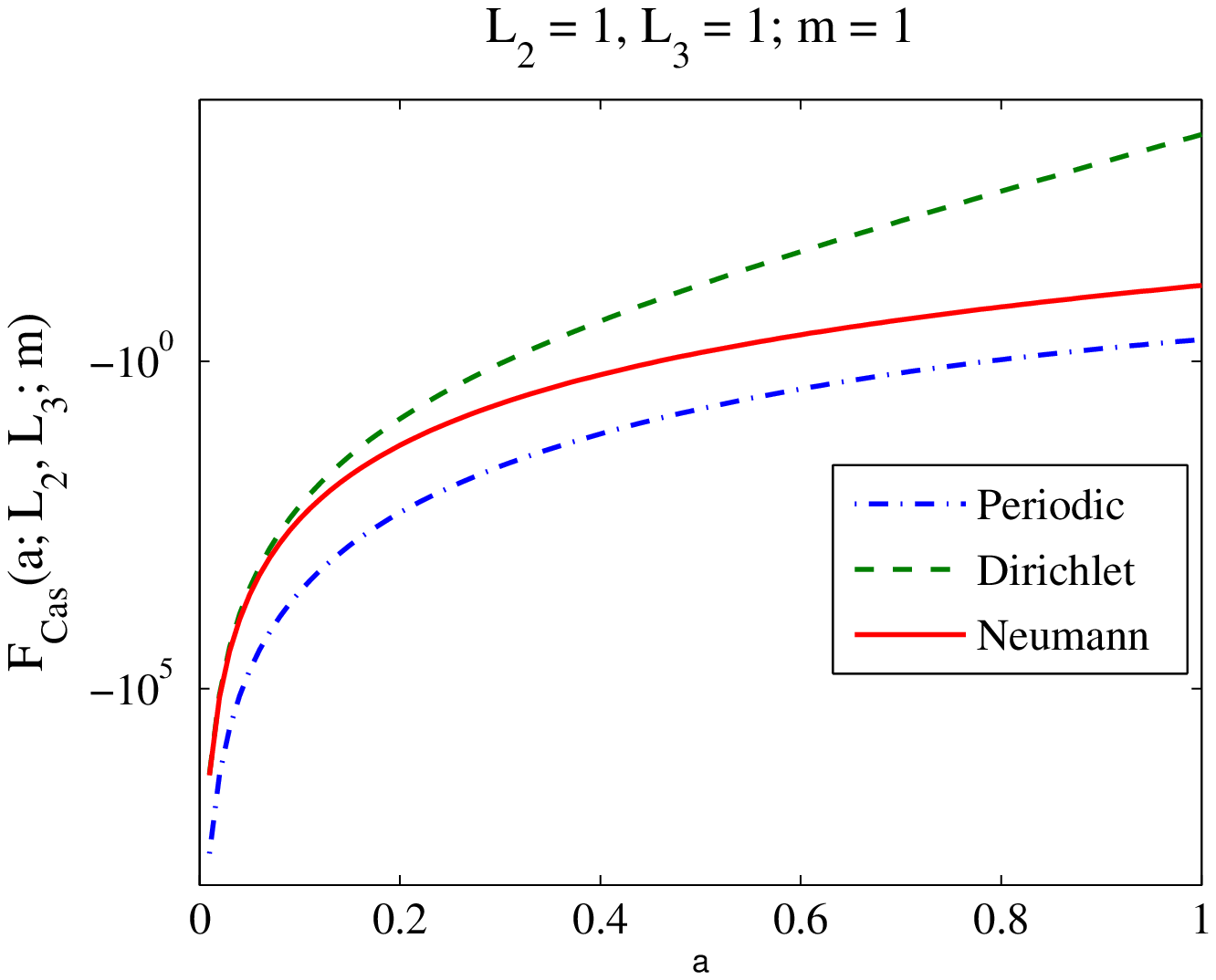}\\\epsfxsize=.49\linewidth
\epsffile{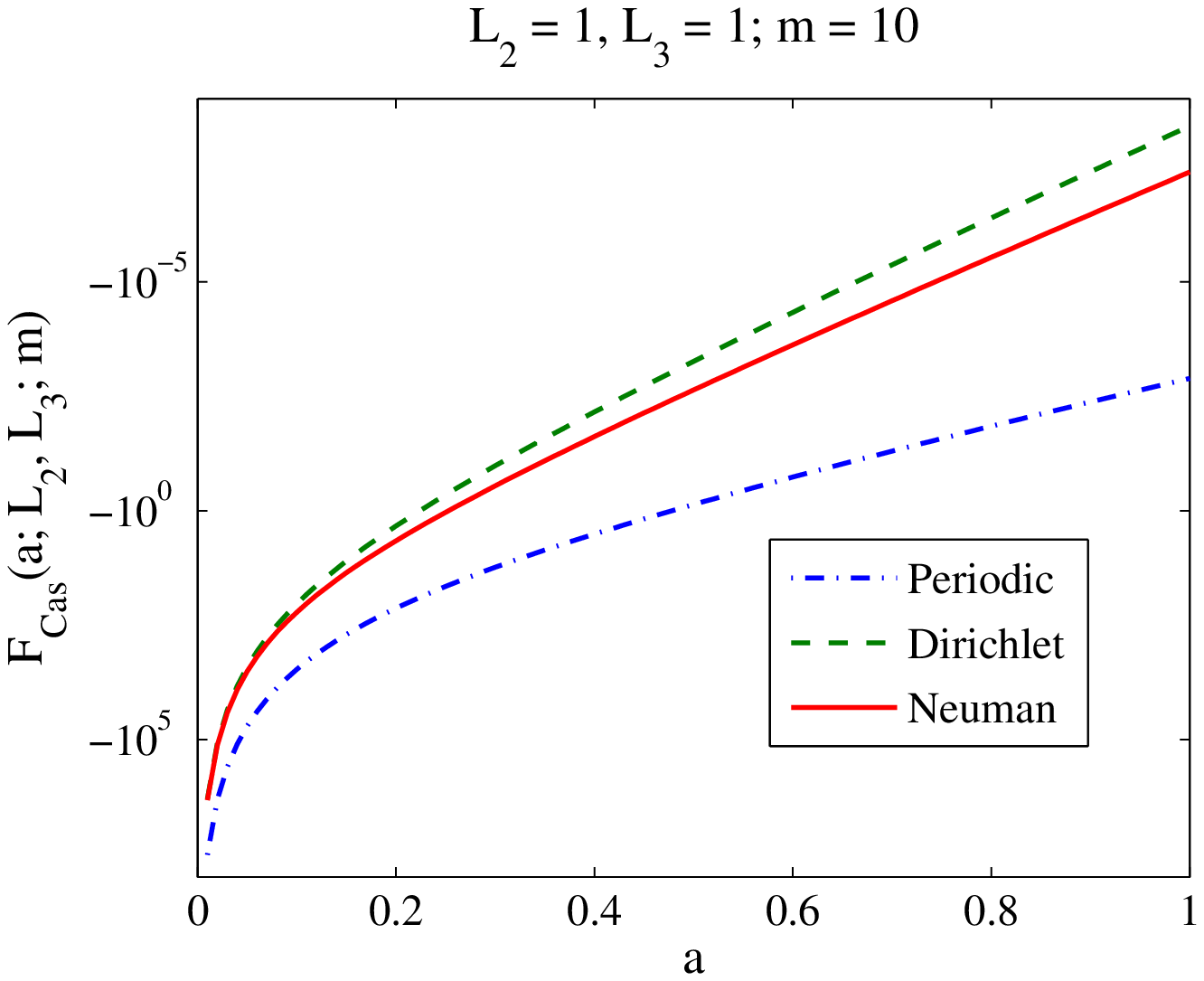}\epsfxsize=.49\linewidth
\epsffile{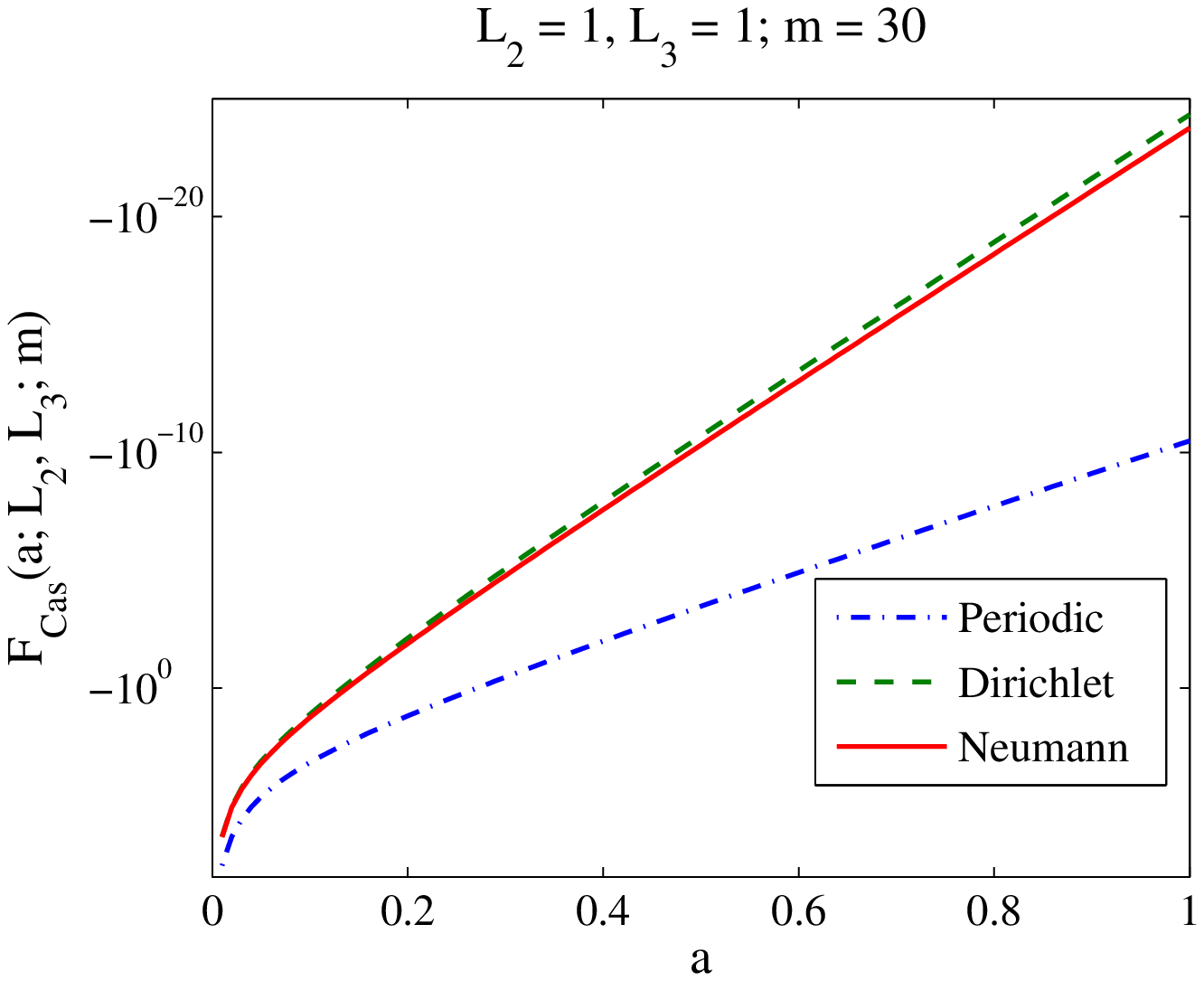}\caption{These figures show the dependence
of the Casimir force $F_{\text{Cas}}^{P/D/N}(a; L_2,L_3; m)$ on $a$
when $L_2=L_3=1$ and $m=0.1, 1, 10, 30$ respectively.}
\end{figure}

 Another advantage of the explicit formulas \eqref{eq6_25_2} and \eqref{eq6_24_5}
 is that since the Bessel function $K_{\nu}(z)$ is positive for all
 $z>0$, Eqs. \eqref{eq6_25_2} and \eqref{eq6_24_5} show immediately that
 \emph{the sign of the Casimir force $F_{\text{Cas}}^{P/D/N}(a;  L_2, L_3;
 m)$ for all the three boundary conditions is negative, and
 therefore they are all attractive forces. } This extends the known
 results about the massless scalar fields \cite{16, 9, 7}. Extension of this result to massive scalar fields in arbitrary dimensions
 will be considered in a future work. Here we would like to remark that as is shown
 in Appendix B, the Casimir force acting on the piston due to the regularized Casimir
 energy of the interior region alone can be either attractive or
 repulsive for any of the boundary conditions. It is the total
 effect of the Casimir force from  both the interior and exterior regions
 that gives a net finite force which is attractive.

For the behavior of the Casimir force when the separation distance
$a$ is small compared to $L_2$ and $L_3$, we apply again the formula
\eqref{eq6_23_1} to the summation over $k_1$ term of
\eqref{eq6_20_10}. The detailed derivation is given in the Appendix C.
Here we present the end results:
\begin{align}\label{eq6_24_1} &F_{\text{Cas}}^{P}(a;  L_2, L_3; m)
\\=&\frac{L_2L_3}{2\pi^2a^2}m^2\sum_{k_1=1}^{\infty}
\frac{1}{k_1^2}K_2(amk_1)-\frac{L_2L_3}{2\pi^2a}m^3\sum_{k_1=1}^{\infty}
\frac{1}{k_1} K_3(amk_1)\nonumber\\
&-\frac{L_2L_3}{4\pi^2}m^2\sum_{(k_2,
k_3)\in\Z^2\setminus\{0\}}\left(\sum_{i=2}^3[k_iL_i]^2\right)^{-1}
K_2\left(m\sqrt{\sum_{i=2}^3[k_iL_i]^2}\right)\nonumber\\
&+\frac{2\sqrt{2\pi}L_2L_3}{a^3} \sum_{k_1=1}^{\infty} \sum_{(k_2,
k_3)\in\Z^2\setminus\{0\}}k_1^2\left(
\frac{m^2+\frac{4\pi^2k_1^2}{a^2}}{\sum_{i=2}^3[k_iL_i]^2}\right)^{1/4}
K_{1/2}\left(\sqrt{\left(m^2+\frac{4\pi^2k_1^2}{a^2}\right)\left(\sum_{i=2}^3[k_iL_i]^2\right)}\right)\nonumber
\end{align}for the periodic bc case; and
\begin{align}\label{eq6_24_2} &F_{\text{Cas}}^{D/N}(a;  L_2, L_3; m)
\\=&\frac{L_2L_3}{8\pi^2a^2}m^2\sum_{k_1=1}^{\infty}
\frac{1}{k_1^2}K_2(2amk_1)-\frac{L_2L_3}{4\pi^2a}m^3\sum_{k_1=1}^{\infty}
\frac{1}{k_1} K_3(2amk_1)\nonumber\\&\mp
\frac{(L_2+L_3)}{8\pi^{3/2}}\frac{m^{3/2}}{a^{3/2}}\sum_{k_1=1}^{\infty}k_1^{-3/2}K_{3/2}(2amk_1)
\pm
\frac{(L_2+L_3)}{4\pi^{3/2}}\frac{m^{5/2}}{a^{1/2}}\sum_{k_1=1}^{\infty}k_1^{-1/2}K_{5/2}(2amk_1)\nonumber\\
&+\frac{1}{8\pi}\frac{m}{a} \sum_{k_1=1}^{\infty}\frac{1}{k_1}
K_1(2amk_1)-\frac{m^2}{4\pi}\sum_{k_1=1}^{\infty}
K_2(2amk_1)\nonumber\\ &\nonumber
\pm\frac{m^{3/2}}{8\pi^{3/2}L_2^{1/2}}\sum_{k_2=1}^{\infty}
k_2^{-3/2}K_{3/2}(2mk_2L_2)
\pm\frac{m^{3/2}}{8\pi^{3/2}L_3^{1/2}}\sum_{k_3=1}^{\infty}
k_3^{-3/2}K_{3/2}(2mk_3L_3)\\ &-\frac{L_2L_3}{16\pi^2}m^2\sum_{(k_2,
k_3)\in\Z^2\setminus\{0\}}\left(\sum_{i=2}^3[k_iL_i]^2\right)^{-1}
K_2\left(2m\sqrt{\sum_{i=2}^3[k_iL_i]^2}\right)\nonumber\\
&\nonumber \mp\frac{\pi
L_2}{2a^3}\sum_{k_1=1}^{\infty}\sum_{k_2=1}^{\infty}
k_1^2K_0\left(2k_2L_2\sqrt{m^2+\frac{\pi^2
k_1^2}{a^2}}\right)\mp\frac{\pi
L_3}{2a^3}\sum_{k_1=1}^{\infty}\sum_{k_3=1}^{\infty}
k_1^2K_0\left(2k_3L_3\sqrt{m^2+\frac{\pi^2 k_1^2}{a^2}}\right)\\
&+\frac{\sqrt{\pi}L_2L_3}{4a^3} \sum_{k_1=1}^{\infty} \sum_{(k_2,
k_3)\in\Z^2\setminus\{0\}}k_1^2\left(
\frac{m^2+\frac{\pi^2k_1^2}{a^2}}{\sum_{i=2}^3[k_iL_i]^2}\right)^{1/4}
K_{1/2}\left(2\sqrt{\left(m^2+\frac{\pi^2k_1^2}{a^2}\right)\left(\sum_{i=2}^3[k_iL_i]^2\right)}\right)\nonumber
\end{align}for the Dirichlet bc and Neumann bc cases. The last term
on the right hand side of \eqref{eq6_24_1} and the last three terms
on the right hand side of \eqref{eq6_24_2} vanish to zero
exponentially fast when $a\rightarrow 0^+$. The third term on the
right hand side of \eqref{eq6_24_1} and the seventh, eighth and
ninth terms on the right hand side of \eqref{eq6_24_2} are
independent of $a$. For the remaining terms, their behaviors as
$a\rightarrow 0^+$ are not obvious. Using the power series expansion
of the Bessel function $K_{\nu}(z)$ naively do not give us the
correct asymptotic behavior due to the summation over $k_1$. In
Appendix C, we derive the correct asymptotic behavior for these
terms when $a\rightarrow 0^+$. The results are: for fixed $m,
L_2,L_3$, as $a\rightarrow 0^+$,
\begin{figure}\centering
\epsfxsize=.49\linewidth
\epsffile{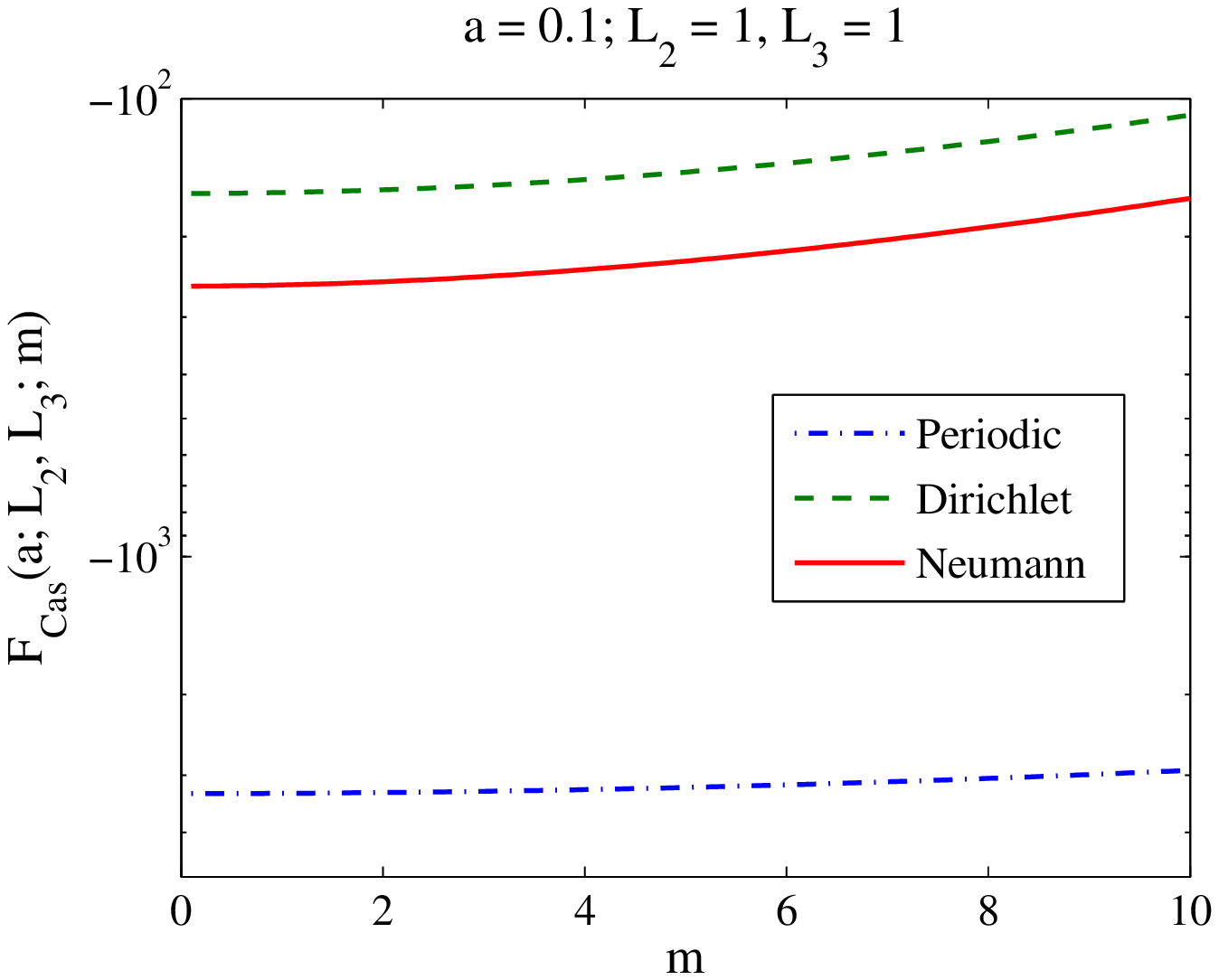}\epsfxsize=.49\linewidth
\epsffile{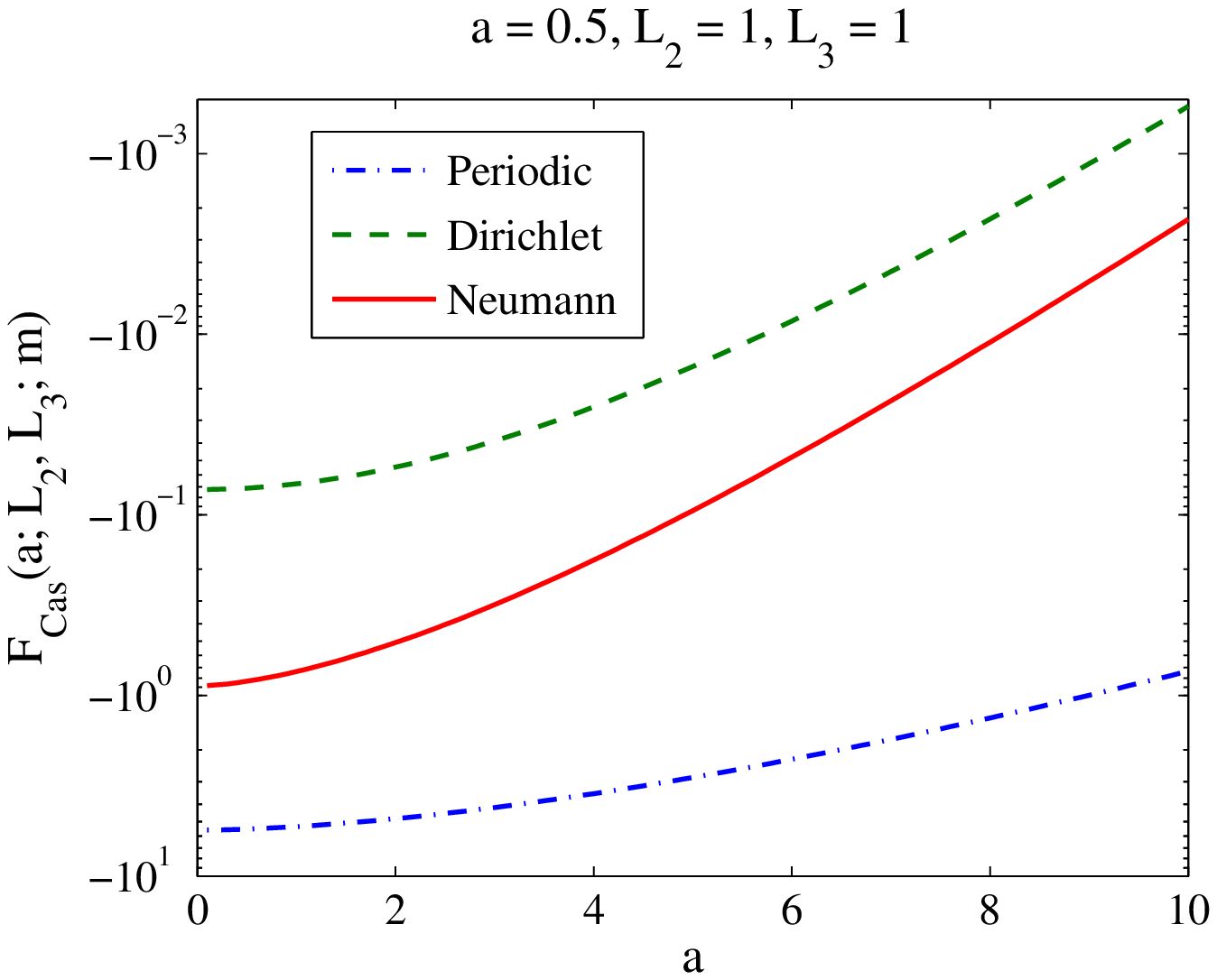}\\\epsfxsize=.49\linewidth
\epsffile{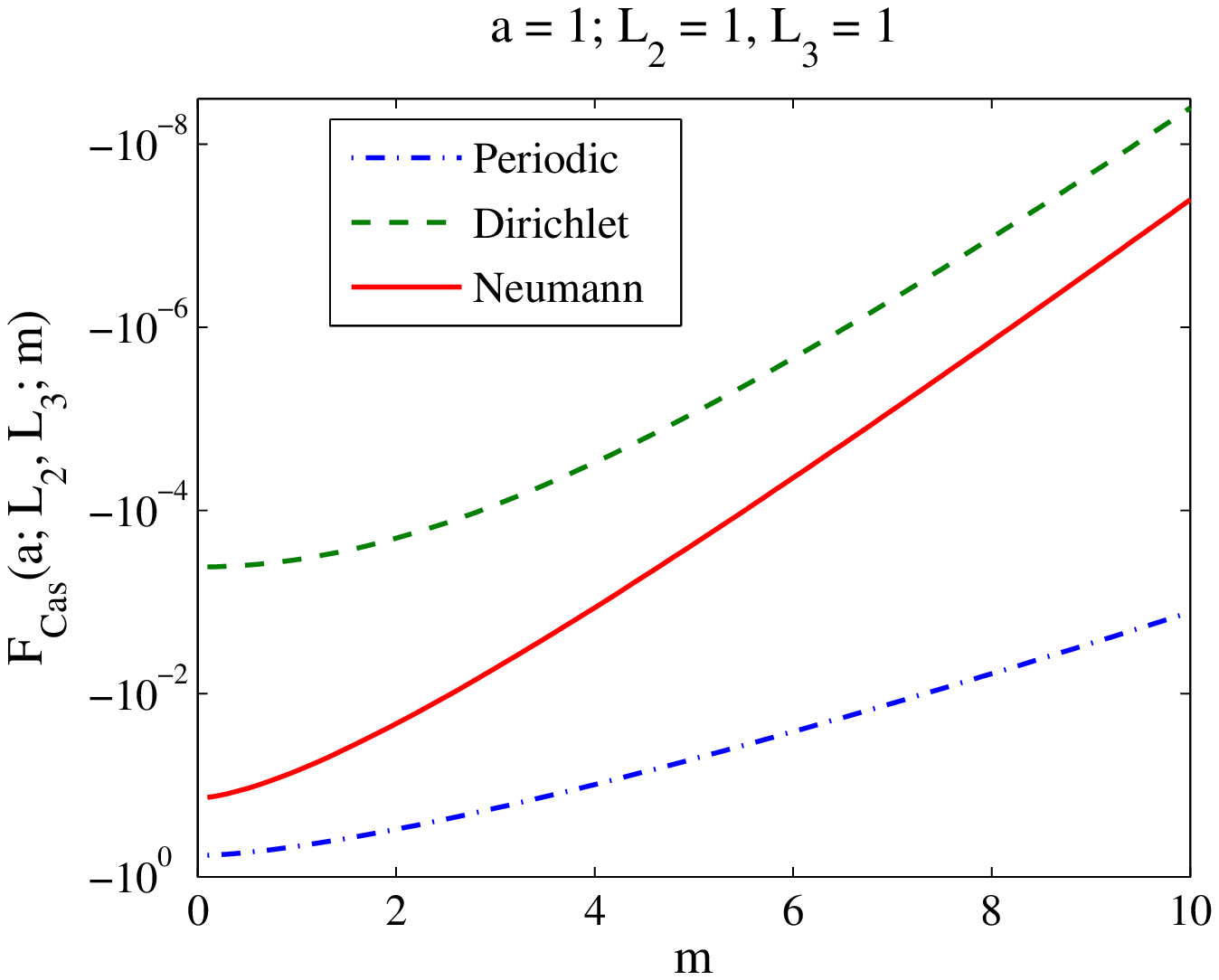}\epsfxsize=.49\linewidth
\epsffile{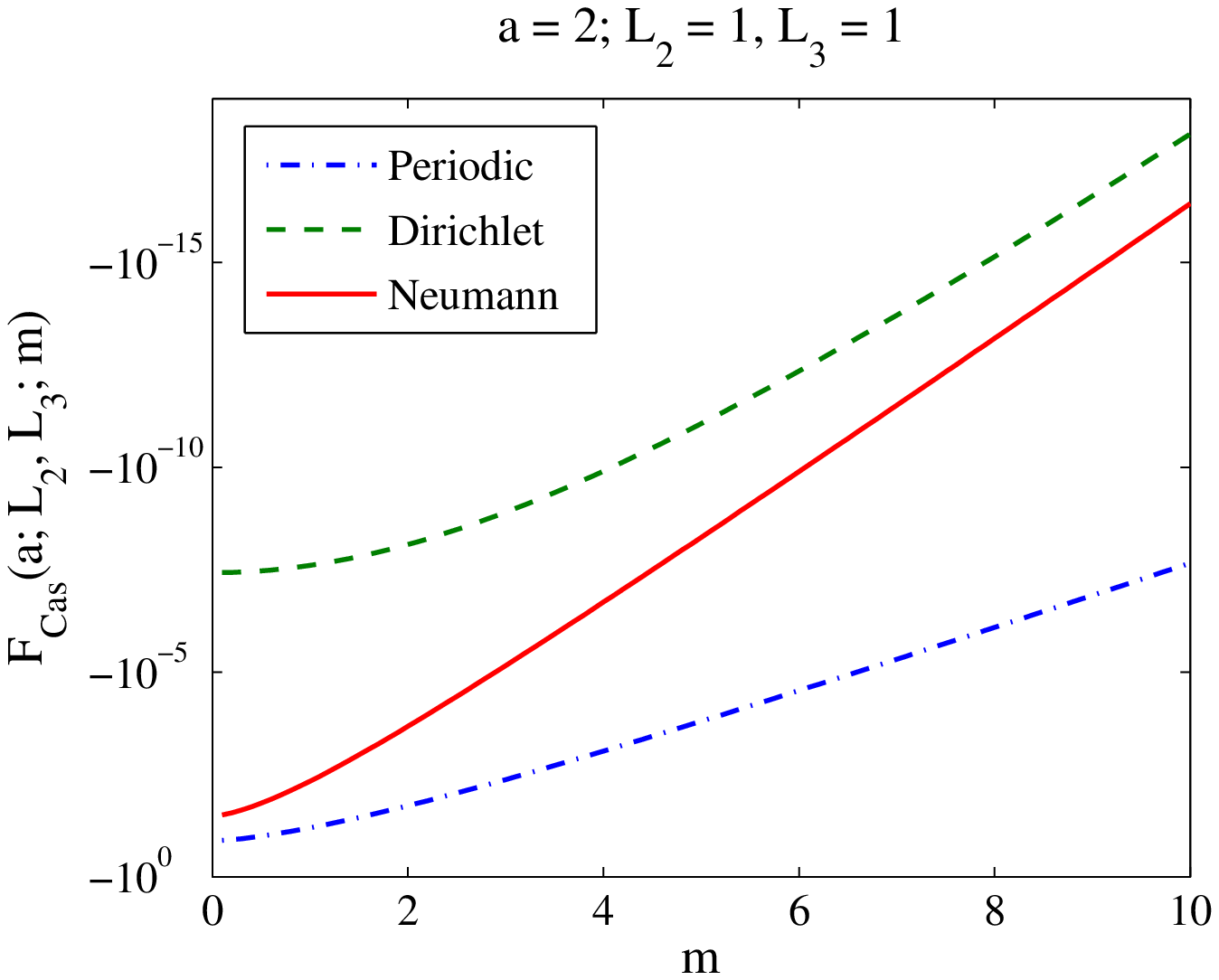}\caption{These figures show the
dependence of the Casimir force $F_{\text{Cas}}^{P/D/N}(a; L_2,L_3;
m)$ on $m$ when $L_2=L_3=1$ and $a=0.1, 0.5, 1, 2$ respectively.}
\end{figure}
\begin{align*}
F_{\text{Cas}}^{P}(a;  L_2, L_3; m)\sim &-\frac{\pi^2L_2L_3}{30
a^4}+\frac{L_2L_3 m^2}{24 a^2}+\frac{\pi^2m^4}{32}L_2L_3\log a+O(a^{0})\\
F_{\text{Cas}}^{D/N}(a;  L_2, L_3; m)\sim &-\frac{\pi^2L_2L_3}{480
a^4}\pm \frac{\zeta(3)(L_2+L_3)}{16\pi
a^3}-\frac{\pi}{96a^2}+\frac{L_2L_3 m^2}{96 a^2}\mp
\frac{m^2(L_2+L_3)}{16\pi a}\\&+\frac{\pi^2m^4}{32}L_2L_3\log
a-\frac{\pi m^2}{16}\log a+O(a^{0}).
\end{align*}In particular, as $a\rightarrow 0^+$, the Casimir force becomes very negative (attractive)
with  leading  term of order $1/a^4$. Notice that this leading term
 \emph{is independent of the mass $m$}.   We also
observe that the first three leading  terms of the Casimir force
$F_{\text{Cas}}^D(a; L_2, L_3;m)$ for Dirichlet bc has exactly the
same form as that of the massless case derived in \cite{9}. In fact,
by taking the limit $m\rightarrow 0^+$ for $F_{\text{Cas}}^D(a; L_2,
L_3;m)$  in \eqref{eq6_24_2}, we obtain the same expression as that
derived in \cite{9} when $a$ is small.

\begin{figure}\centering
\epsfxsize=.49\linewidth
\epsffile{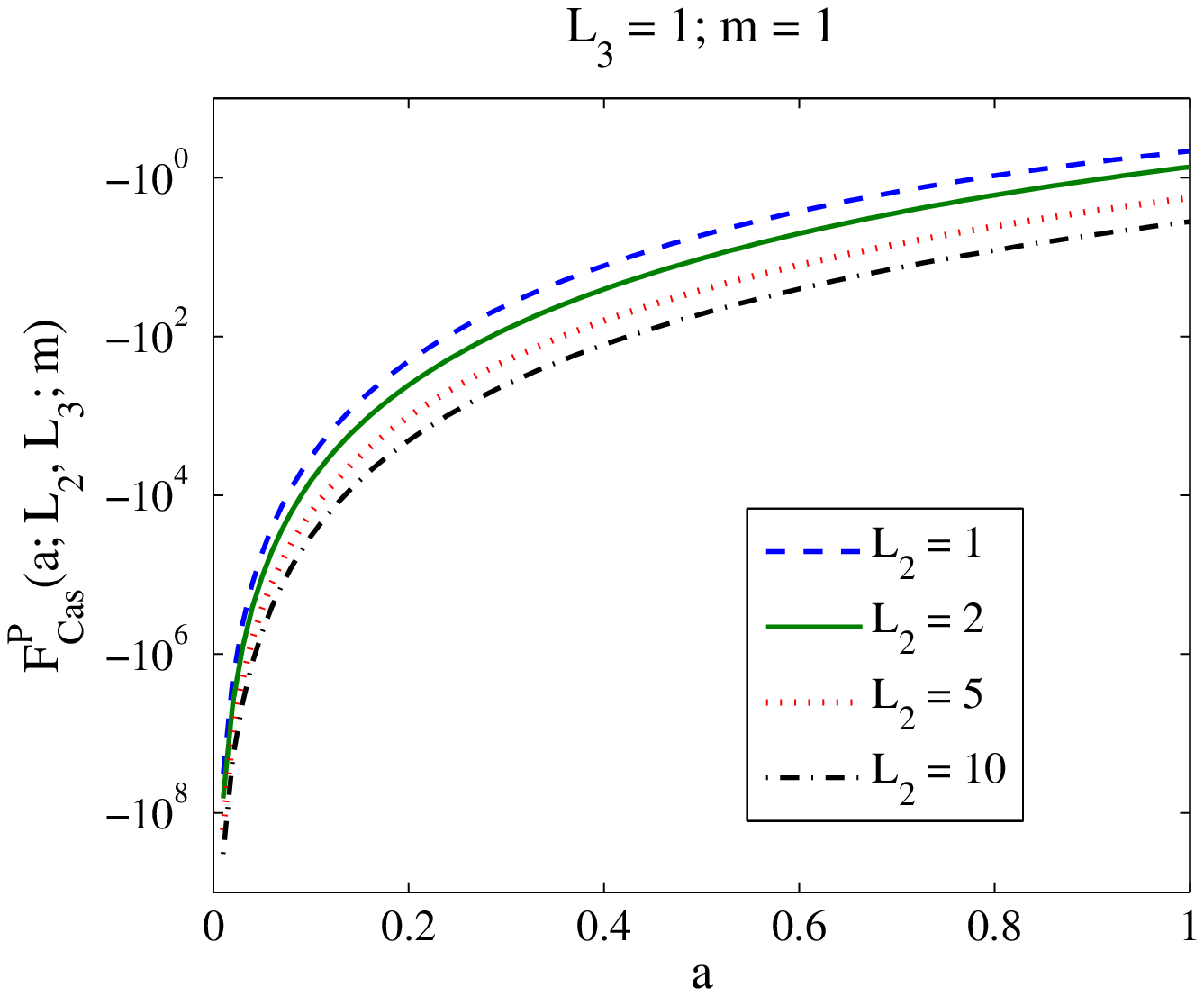}\epsfxsize=.49\linewidth
\epsffile{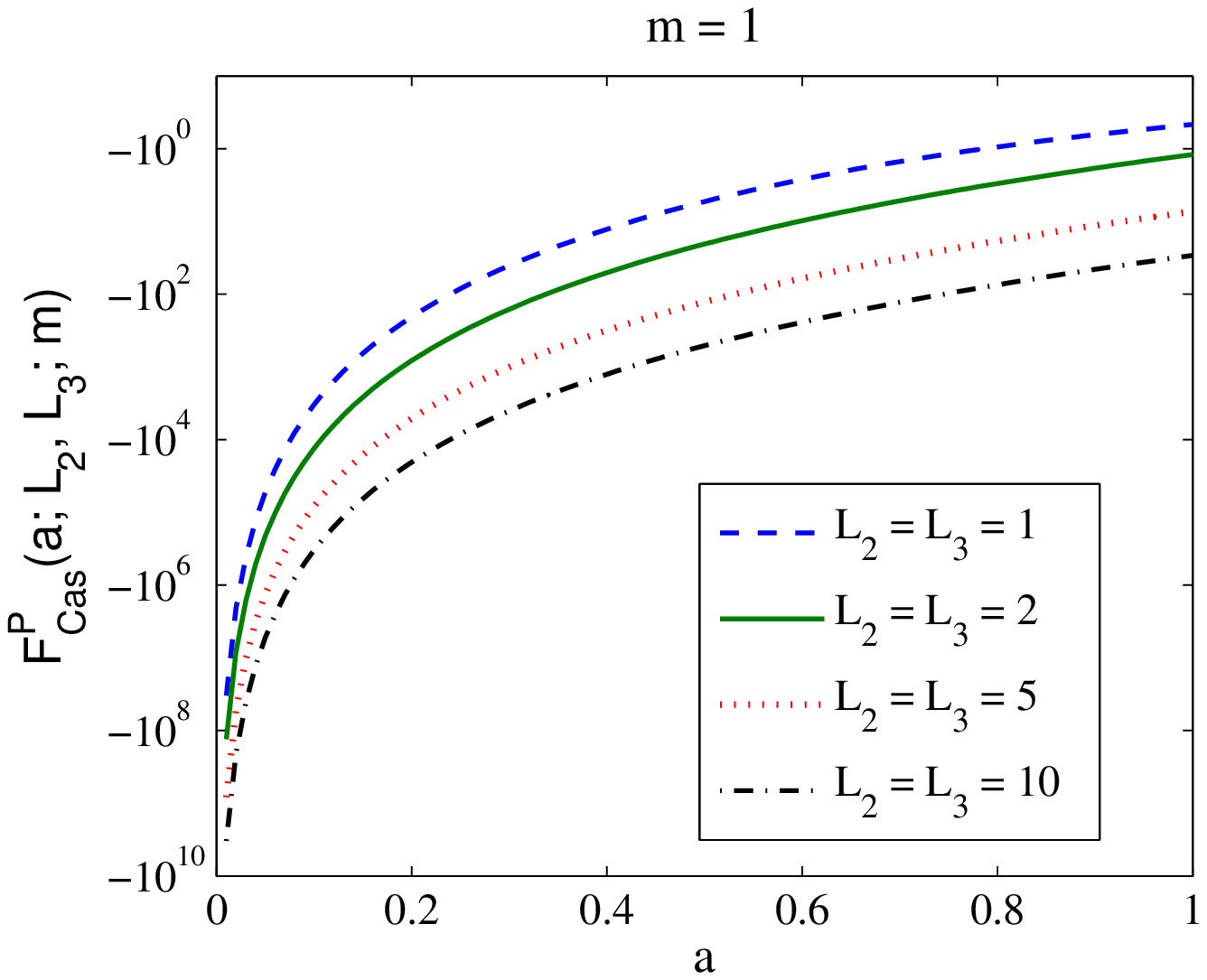}\caption{These figures show
the dependence of the Casimir force $F_{\text{Cas}}^{P}(a; L_2,L_3;
m)$ (periodic bc) on $a$ when $L_2$ and $L_3$ are varied. On the
left hand side, $L_3$ is fixed to equal to 1 and $L_2=1, 2, 5, 10$.
On the right hand side, $L_2$ is set to equal to $L_3$.}
\end{figure}

\begin{figure}\centering
\epsfxsize=.49\linewidth
\epsffile{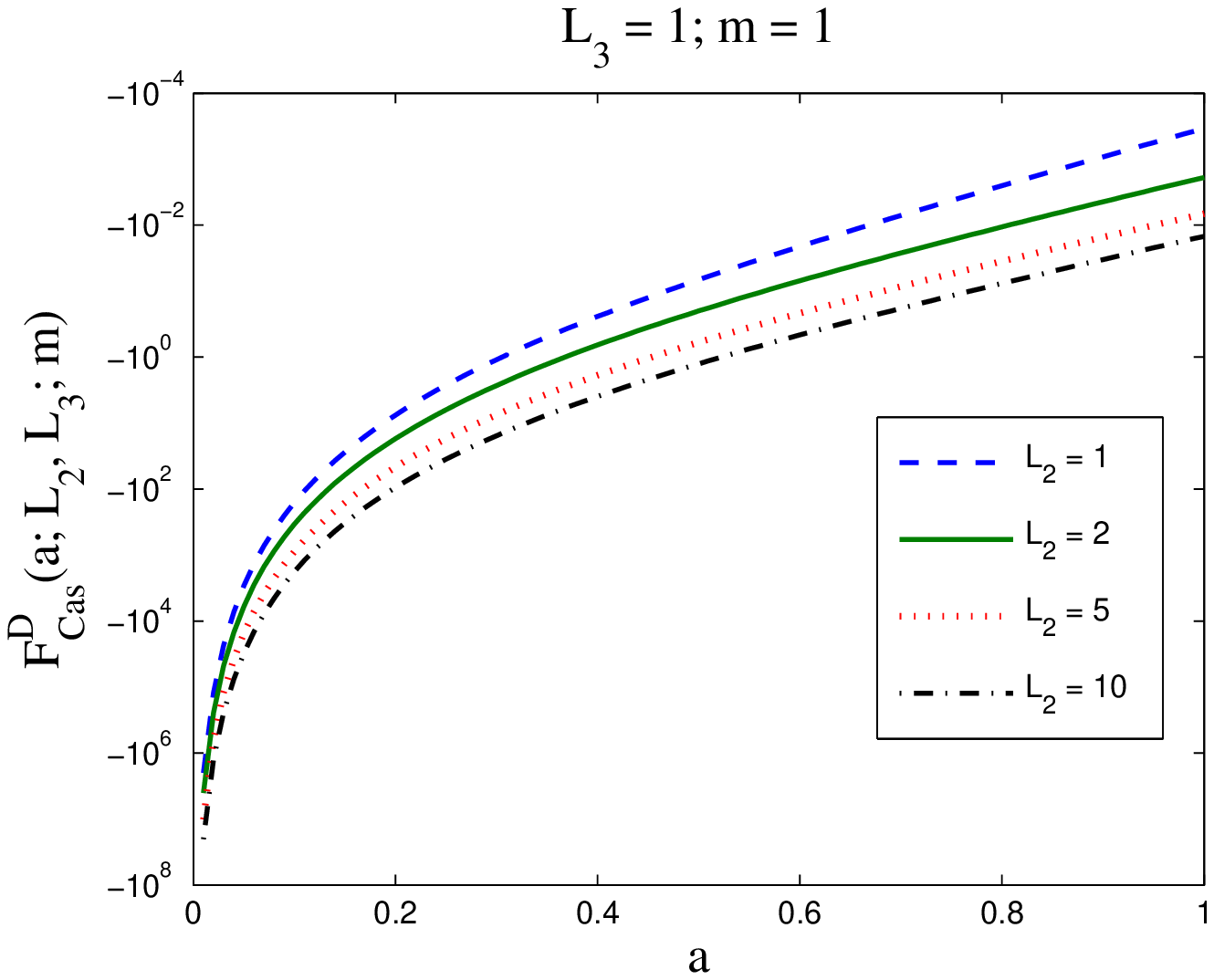}\epsfxsize=.49\linewidth
\epsffile{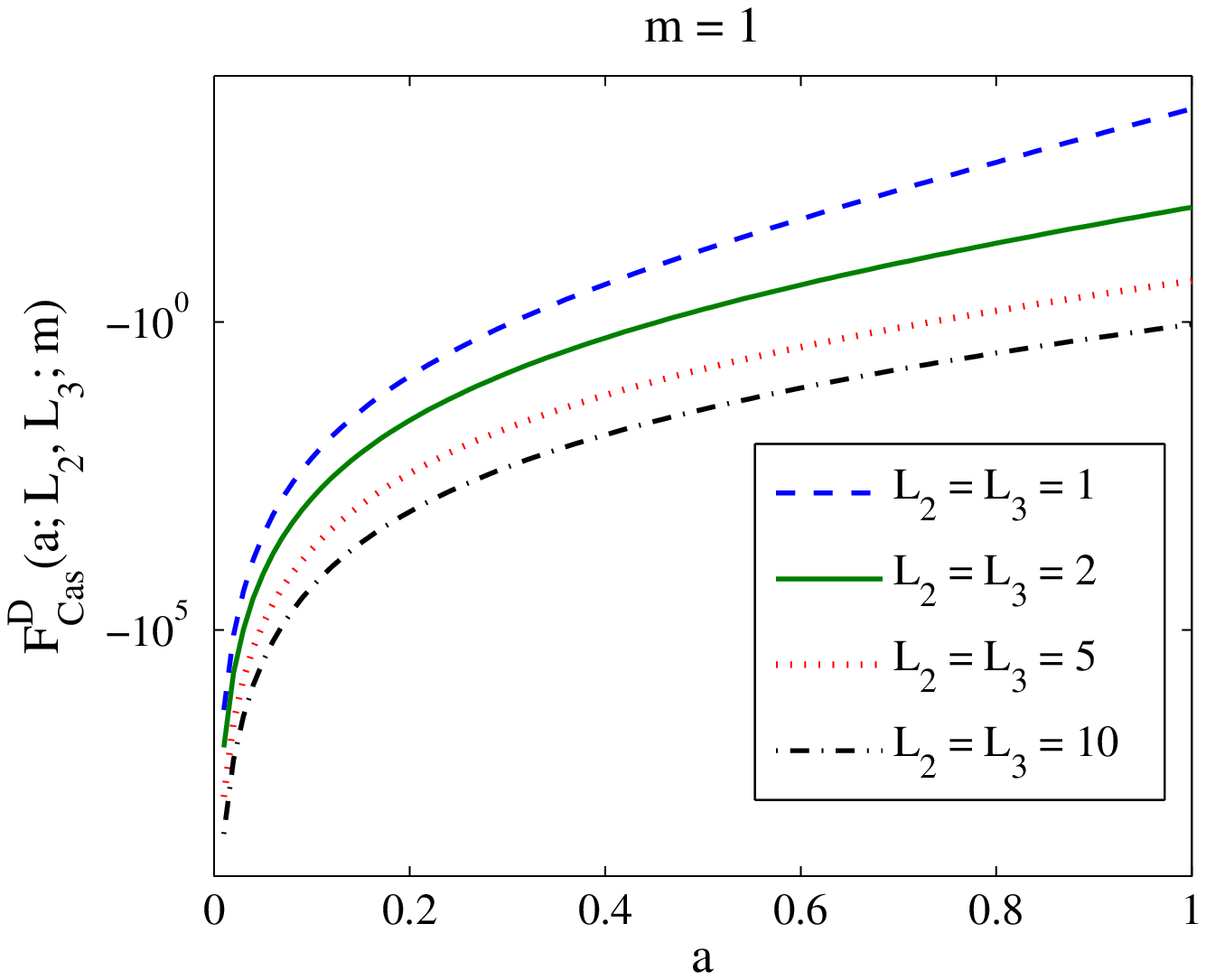}\caption{These figures show
the dependence of the Casimir force $F_{\text{Cas}}^{D}(a; L_2,L_3;
m)$ (Dirichlet bc) on $a$ when $L_2$ and $L_3$ are varied. On the
left hand side, $L_3$ is fixed to equal to 1 and $L_2=1, 2, 5, 10$.
On the right hand side, $L_2$ is set to equal to $L_3$.}
\end{figure}

\begin{figure}\centering
\epsfxsize=.49\linewidth
\epsffile{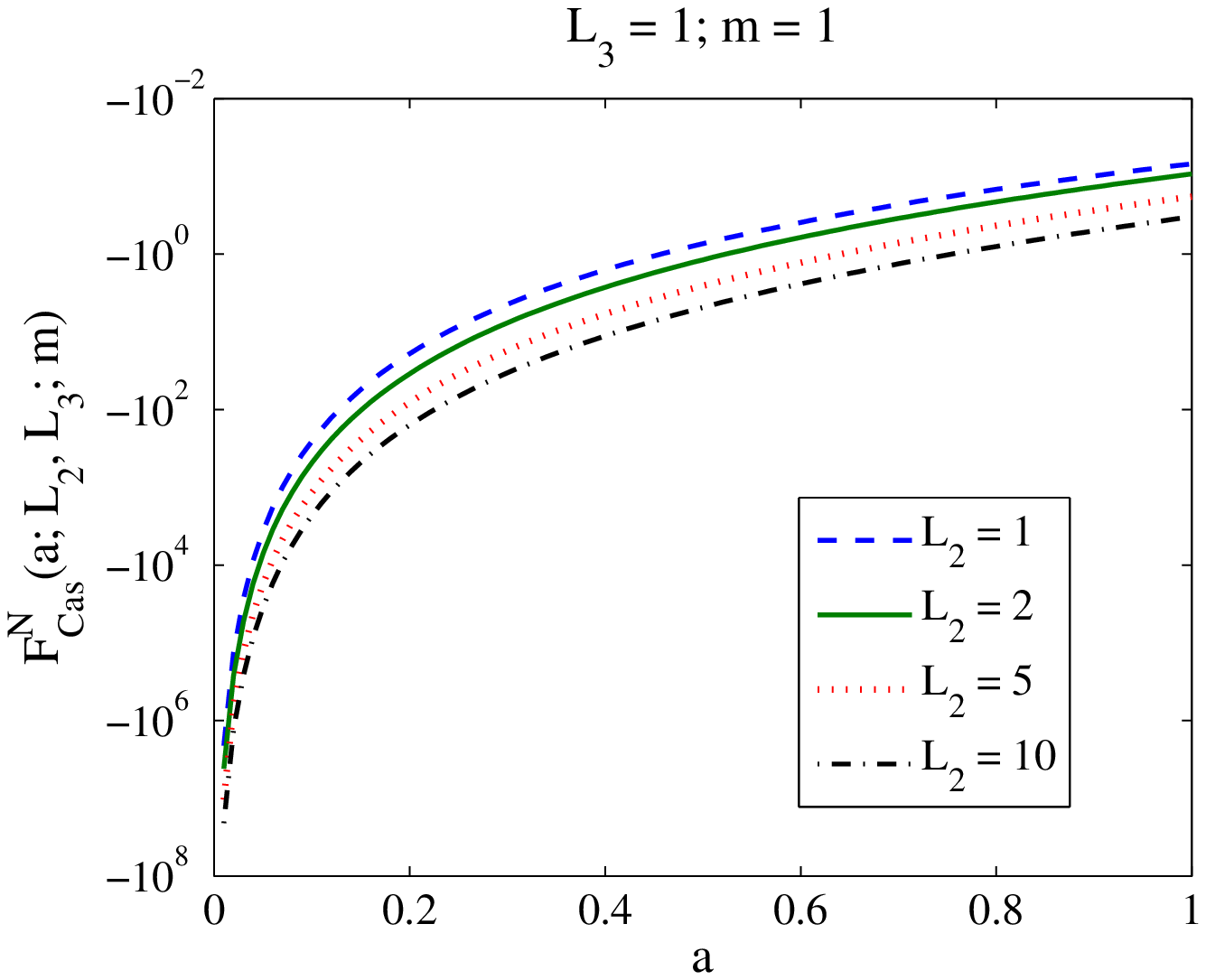}\epsfxsize=.49\linewidth
\epsffile{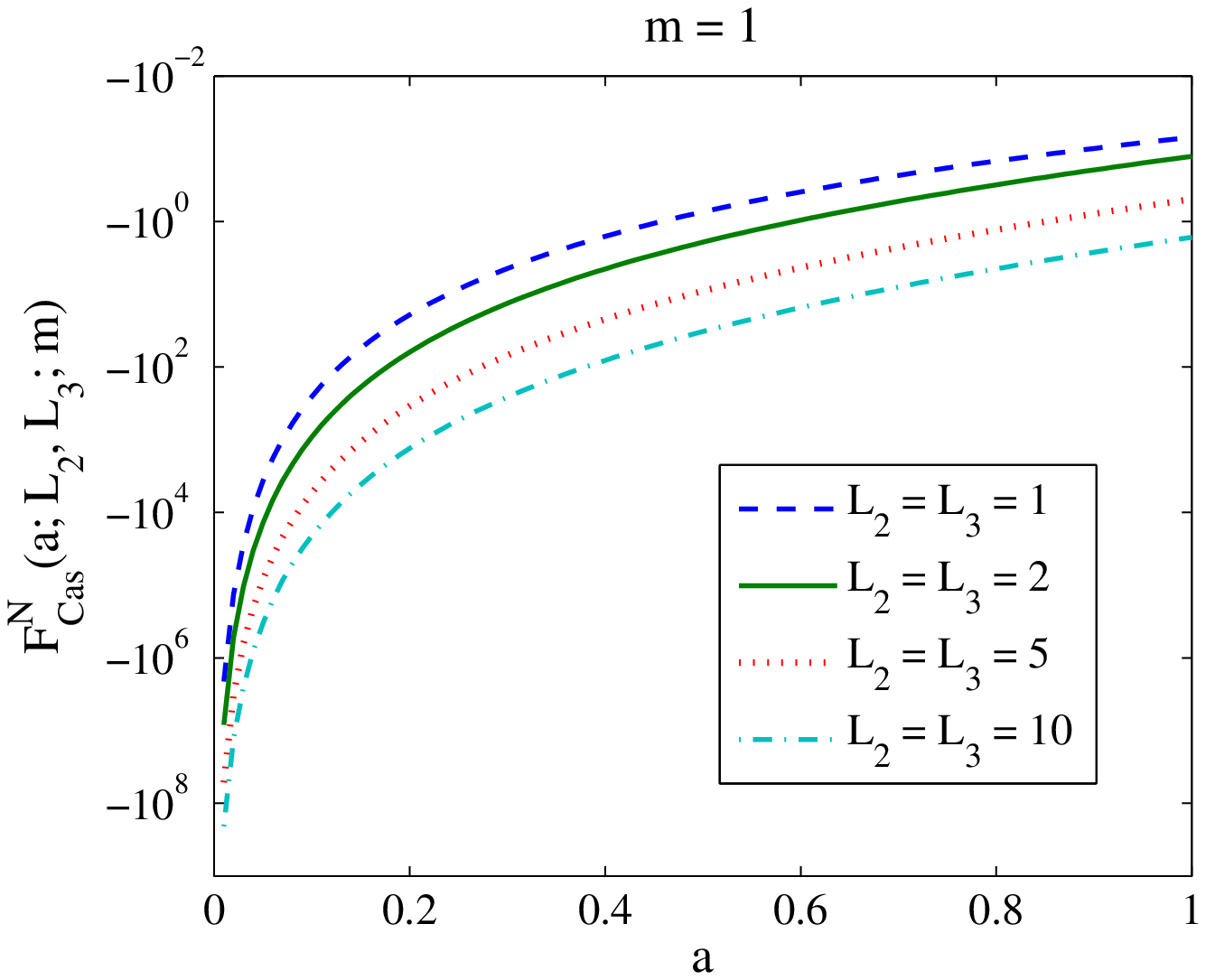}\caption{These figures show
the dependence of the Casimir force $F_{\text{Cas}}^{N}(a; L_2,L_3;
m)$ (Neumann bc) on $a$ when $L_2$ and $L_3$ are varied. On the left
hand side, $L_3$ is fixed to equal to 1 and $L_2=1, 2, 5, 10$. On
the right hand side, $L_2$ is set to equal to $L_3$.}
\end{figure}

From the above discussion, we conclude that the effect of mass  is insignificant
when $a$ is small, but with nonzero mass, the Casimir force will be drastically reduced.

We would also like to mention that the expressions \eqref{eq6_24_1}
and \eqref{eq6_24_2} can also be  used to analyze the behavior of
the Casimir force when $L_2$ and $L_3$ are large (parallel plate
geometry). In this case, we find that the Casimir pressure (force
per unit area) acting on one of the plates is
\begin{align*}
P_{\text{Cas}}^P(a; m) := &\lim_{L_2, L_3\rightarrow \infty}
\frac{F_{\text{Cas}}(a; L_2,L_3; m)}{L_2L_3}
\\=&\frac{m^2}{2\pi^2a^2}\sum_{k_1=1}^{\infty}
\frac{1}{k_1^2}K_2(amk_1)-\frac{m^3}{2\pi^2a}\sum_{k_1=1}^{\infty}
\frac{1}{k_1} K_3(amk_1),\\
P_{\text{Cas}}^{D/N}(a;
m)=&\frac{m^2}{8\pi^2a^2}\sum_{k_1=1}^{\infty}
\frac{1}{k_1^2}K_2(2amk_1)-\frac{m^3}{4\pi^2a} \sum_{k_1=1}^{\infty}
\frac{1}{k_1} K_3(2amk_1)\nonumber.
\end{align*}
In the massless limit $m\rightarrow 0^+$, these become
$P_{\text{Cas}}^P(a; 0)=-\pi^2/(30a^4)$ and $P_{\text{Cas}}^{D/N}(a;
0)=-\pi^2/(480 a^4)$ respectively.

In Figure 2, 3, 4, 5, 6, 7 we show graphically the behavior of the
Casimir force $F_{\text{Cas}}^{P/D/N}(a; L_2, L_3; m)$.
 From \eqref{eq6_25_2} and
\eqref{eq6_24_5}, it is easy to verify that
\begin{align*}
F_{\text{Cas}}^{P/D/N}\left(\lambda a; \lambda L_2, \lambda L_3;
\frac{m}{\lambda}\right)=\lambda^{-2} F_{\text{Cas}}^{P/D/N}(a; L_2,
L_3; m).
\end{align*}Therefore, we have the freedom of fixing one of the
variables which does not affect the shape (up to scaling) of
the graphs. When one is not concerned with the dependence of the
Casimir force on mass, one can let $m=1$. From Figs. 2, 4, 5,
6, we see that fixing $m, L_2$ and $L_3$, the Casimir force always
increase (as a function of $a$) from $-\infty$ (when $a\rightarrow
0^+$) to $0$ (when $a\rightarrow -\infty$). In fact, by
differentiating \eqref{eq6_25_2} and \eqref{eq6_24_5} again with
respect to $a$ and using \eqref{eq6_25_3}, we can conclude that the
Casimir force $F_{\text{Cas}}^{P/D/N}(a; L_2, L_3; m)$ is always an
increasing function of $a$, and therefore the small $a$ and large
$a$ behaviors show that it always increases from $-\infty$ (when
$a\rightarrow 0^+$) to $0$ (when $a\rightarrow -\infty$).

Fig. 3 shows the dependence of the Casimir force on mass $m$. We see
that the mass effect is more significant when $a$ is large. On the
other hand, we will tend to make conclusion that the Casimir force
is an increasing function of $m$ from Fig. 3. However, Fig. 7
invalidates this conjecture.

\begin{figure}\centering
\epsfxsize=.49\linewidth
\epsffile{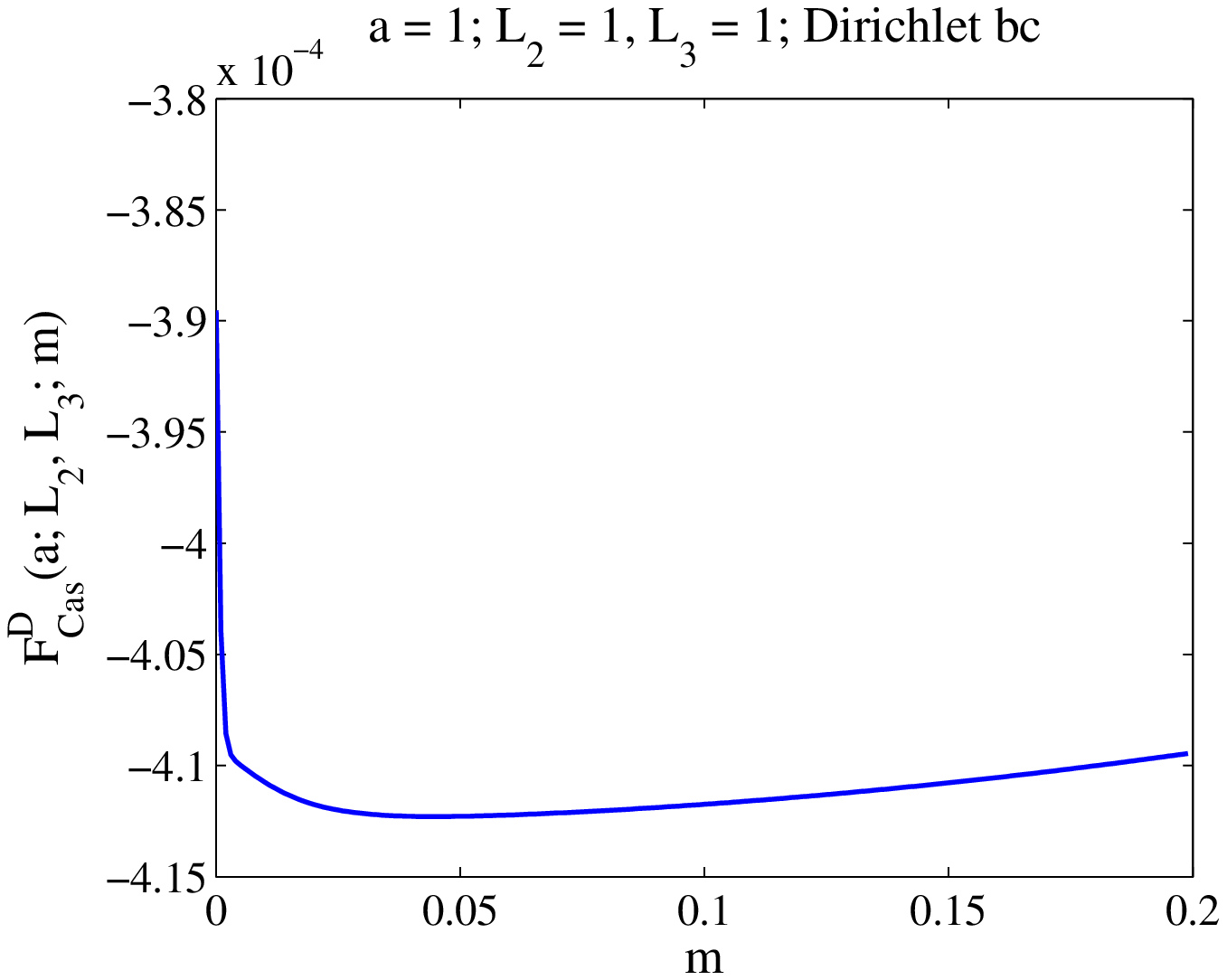}\epsfxsize=.49\linewidth
\epsffile{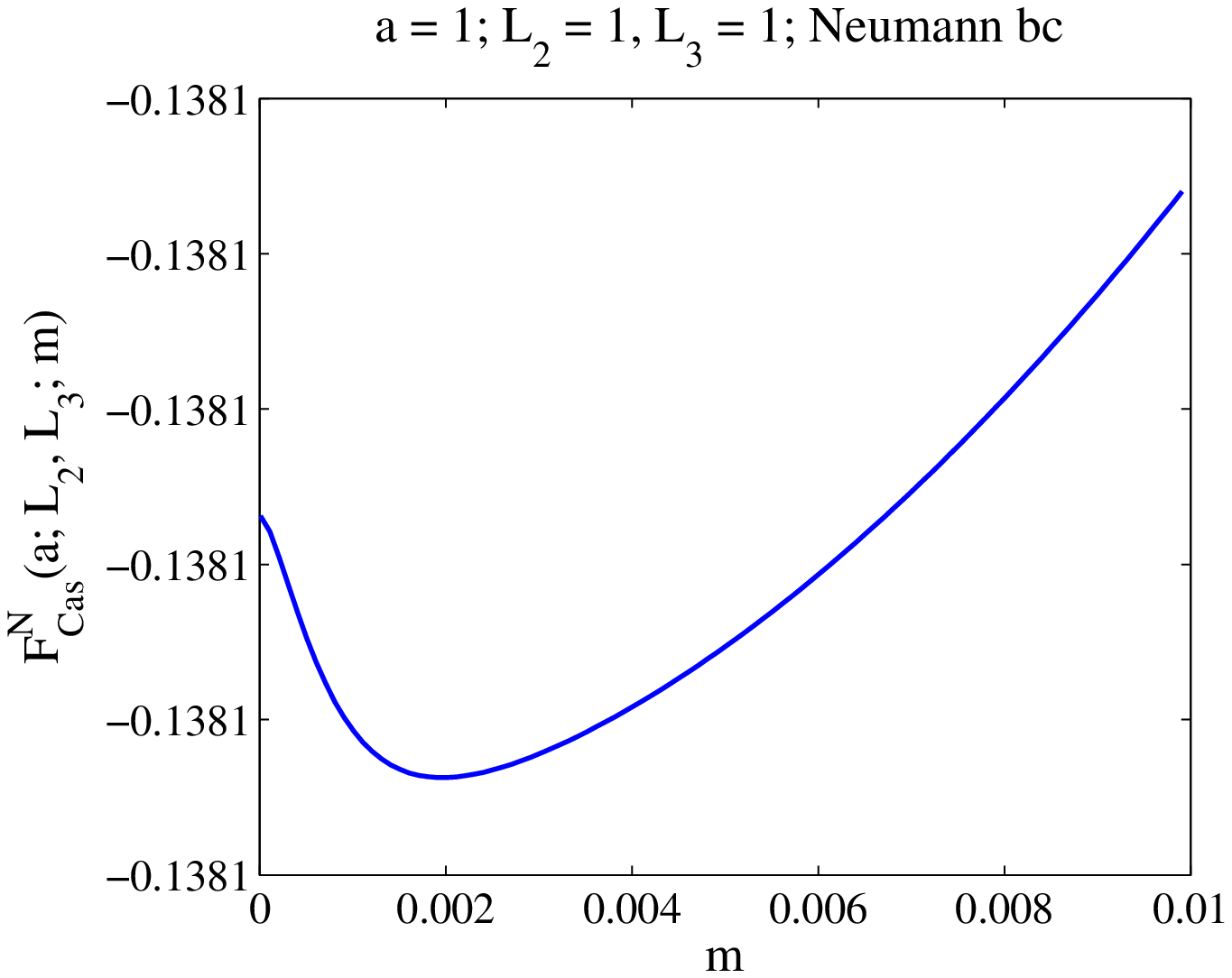}\caption{These figures show that
the  Casimir force $F_{\text{Cas}}(a; L_2,L_3; m)$ are not monotonic
 function of mass.}
\end{figure}

In Figure 8, we demonstrate the Casimir force on the
 piston, the regularized Casimir force and reduced regularized Casimir force (see Appendix B) due to the
 interior and exterior regions on a single graph, for the periodic
 bc and Dirichlet bc cases. The graphs show that the Casimir force
 due to the interior region and exterior exterior region can be
 either attractive or repulsive. But the net effect on the Casimir
 piston is attractive.
\begin{figure}\centering
\epsfxsize=.49\linewidth
\epsffile{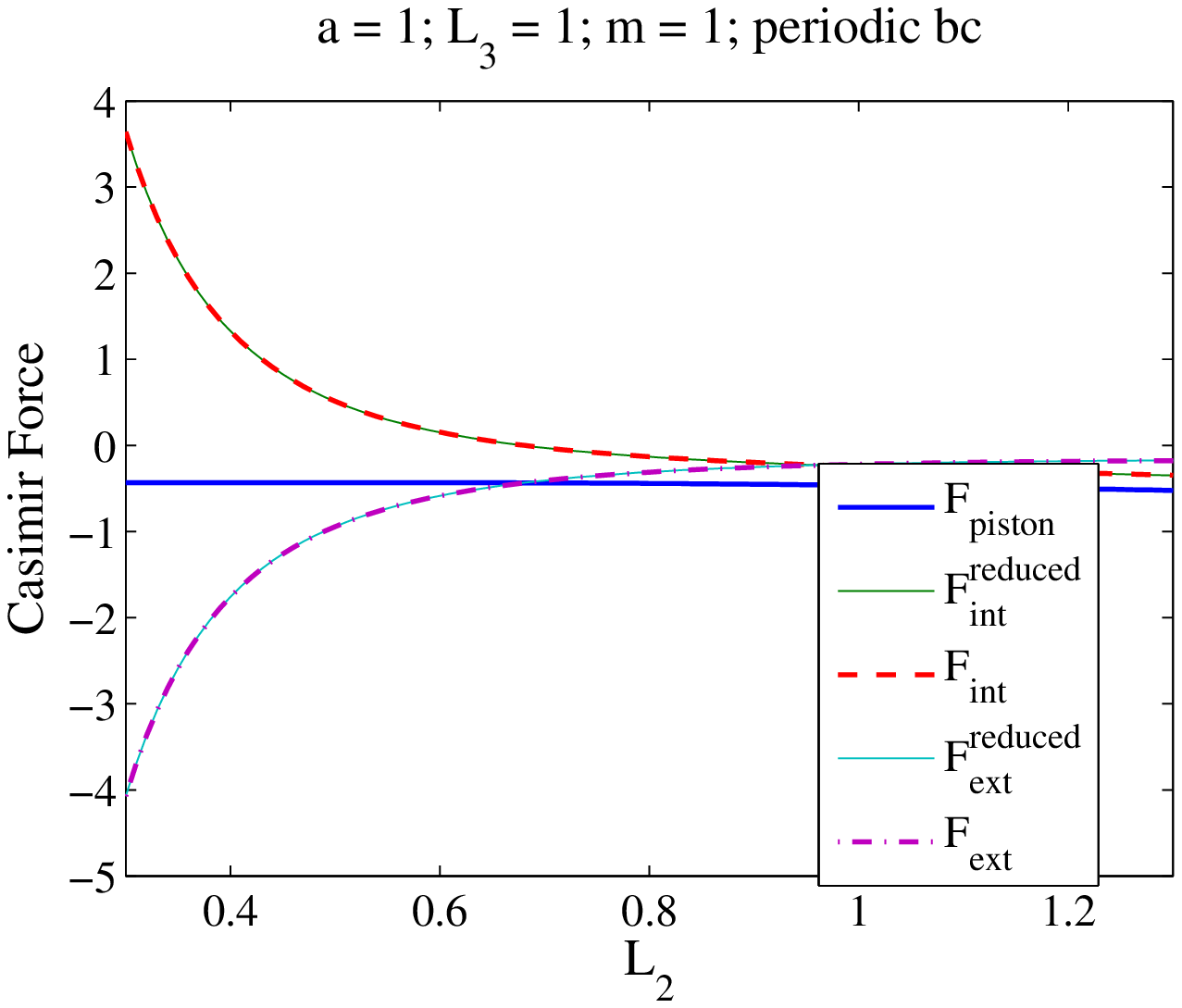}\epsfxsize=.49\linewidth
\epsffile{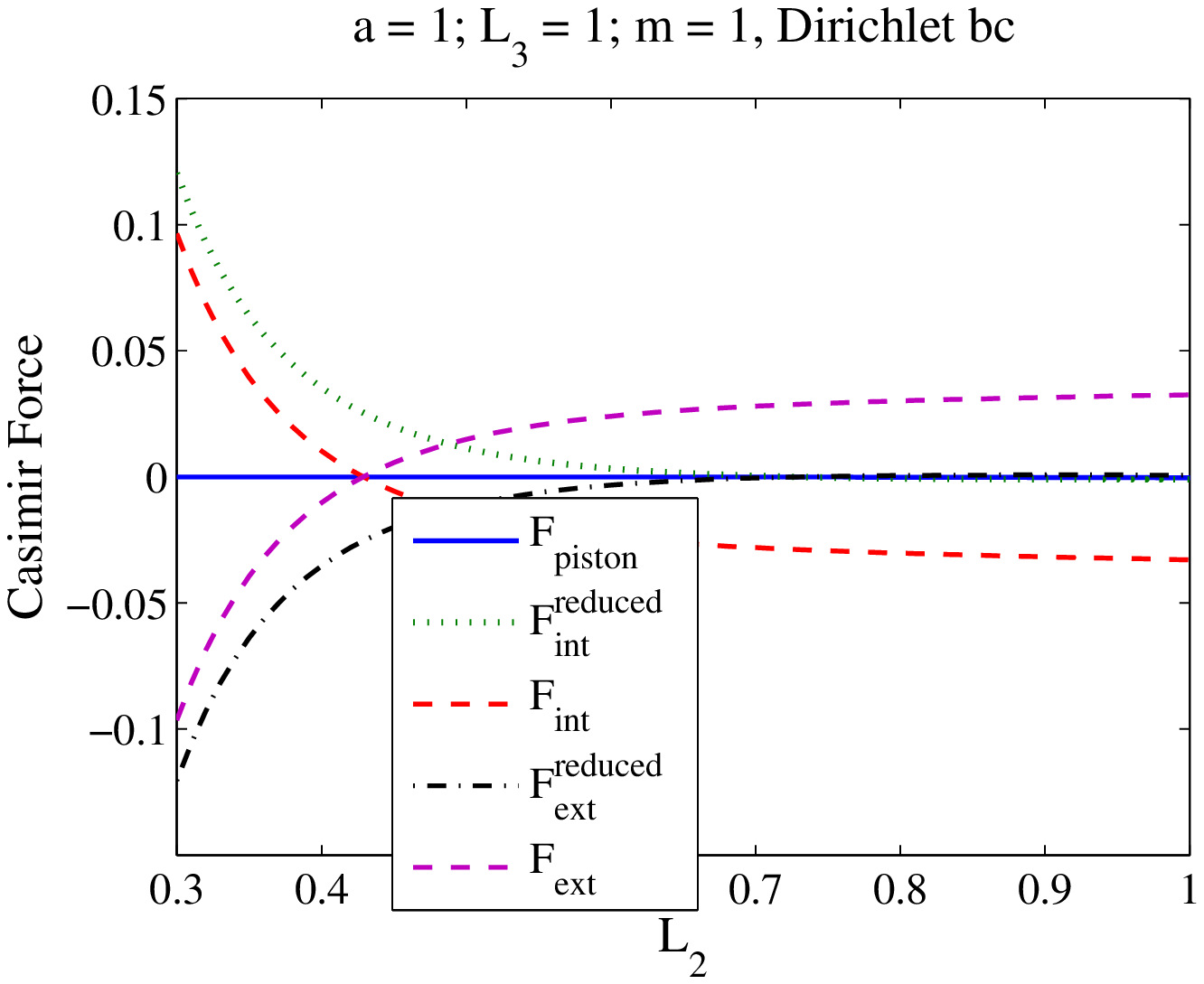}\caption{These figures show that
despite the fact that the Casimir force acting on the piston is
always attractive, the regularized Casimir force due to the interior
region and exterior regions can be either attractive or repulsive.}
\end{figure}

In the above discussion, we are considering the case where the limit
$L_1 \rightarrow \infty$ has been  taken. Now we briefly discuss the
case where the limit $L_1\rightarrow \infty$ is not taken, i.e. the
piston is assumed to be placed in a fixed closed rectangular box. To
distinguish with the cases above, we denote the resulting Casimir
force on the piston as $\tilde{F}_{\text{Cas}}^{P/D/N}(a; L_2, L_3;
m)$. From the discussion in the beginning of this section,
$\tilde{F}_{\text{Cas}}^{P/D/N}(a; L_2, L_3; m)$ is given by the
right hand side of \eqref{eq6_27_2} and \eqref{eq6_27_3} without the
limit $\lim_{L_1\rightarrow \infty}$. Therefore, in terms of the
Casimir force $F_{\text{Cas}}^{P/D/N}(a; L_2, L_3; m)$ computed for
the case $L_1\rightarrow \infty$, we have
\begin{align*}
\tilde{F}_{\text{Cas}}^{P/D/N}(a; L_2, L_3;
m)=F_{\text{Cas}}^{P/D/N}(a; L_2, L_3;
m)-F_{\text{Cas}}^{P/D/N}(L_1-a; L_2, L_3; m)
\end{align*}Obviously $\tilde{F}_{\text{Cas}}^{P/D/N}(a; L_2, L_3;
m)=0$ when $a=L_1/2$, i.e. the piston is in an equilibrium position
when it is placed in the middle of the box. However, if it is placed
closer to the left end, i.e. $a<L_1-a$, then since
$F_{\text{Cas}}^{P/D/N}(a; L_2, L_3; m)$ is more negative than
$F_{\text{Cas}}^{P/D/N}(L_1-a; L_2, L_3; m)$, we find a net
attractive force pulling the piston to the left end. In other
words, the Casimir force always tends to move the piston to the
nearest wall. In nanotechnology, this is known as the stiction
effect and may have undesirable consequences to the functionality of
nano--devices.
\section{Conclusion}

  We have carried out the study of the Casimir effect for massive scalar
fields in three dimensions subject to different boundary conditions
in the piston setting. The methods we used here can be easily
generalized to show that for  pistons with arbitrary cross sections
in any dimensions, the Casimir force acting on the piston  is always
divergence free. For the cases   considered here, the Casimir force
is found to be always attractive. It would be interesting to
investigate the possible repulsive Casimir force due to  massive
 fields in the piston  setting. Another thing we have not addressed
 here is the thermal effect, which we plan to carry out   in  the future.

\vspace{1cm} \noindent \textbf{Acknowledgement}\; The authors would
like to thank Malaysian Academy of Sciences, Ministry of Science,
Technology  and Innovation for funding this project under the
Scientific Advancement Fund Allocation (SAGA) Ref. No P96c.

\appendix
\section{Cut-off dependent Casimir energy for massive scalar fields
in rectangular cavities} The cut-off dependent Casimir energy
$E_{\text{Cas}}^{P}(\lambda; L_1, \ldots, L_d;m)$ \eqref{eq6_19_3}
for a massive scalar field with periodic bc inside a $d$-dimensional
rectangular cavity can be written as
\begin{align}\label{eq6_19_5} E_{\text{Cas}}^{P}(\lambda; L_1, \ldots,
L_d;m) =-\frac{1}{2}\frac{\pa}{\pa \lambda}K(\lambda; L_1,\ldots,
L_d;m),
\end{align} where
$$K(\lambda; L_1, \ldots, L_d;m)=\sum_{\mathbf{k}\in \Z^d} e^{-\lambda
\omega_{\mathbf{k}}^P}.$$ For a massless ($m=0$) scalar field, the
sum \eqref{eq6_19_5} has been computed in \cite{6} using
Euler-Maclaurin summation formula. Here we want to compute this kind
of sum by a different method, making full advantage of the analytic
structure of Epstein zeta function. Using inverse Mellin transform
of exponential function
\begin{align*}
e^{-z} =\frac{1}{2\pi i}\int_{c_0-i\infty}^{c_0+i\infty} \Gamma(w)
z^{-w} dw,\hspace{1cm} c_0>0,
\end{align*}we find that
\begin{align}\label{eq6_19_4}
K(\lambda; L_1, \ldots,  L_d)=\frac{1}{2\pi
i}\int_{c_0-i\infty}^{c_0+i\infty} \lambda^{-w}\Gamma(w)Z_{d}
\left(\frac{w}{2}; \frac{2\pi}{L_1},
\ldots,\frac{2\pi}{L_d};m\right) dw, \hspace{1cm}c_0>d,
\end{align}where $Z_d(s; a_1, \ldots, a_d; m)$ is the inhomogeneous
Epstein zeta function defined by the series
\begin{align*}
Z_d(s; a_1,\ldots, a_d; m) =\sum_{\mathbf{k}\in\Z^d}
\left(\sum_{i=1}^d\left[a_ik_i\right]^2 +m^2\right)^{-s}
\end{align*}when $\text{Re}\; s> d/2$. $Z_{d}(s; a_1, \ldots, a_d;m)$ has a meromorphic continuation to
$\C$ given by (see e.g. \cite{1, 2, 3, 4})
\begin{align}\label{eq6_26_1}
&Z_{d} (s;a_1,\ldots, a_d; m)
=\frac{\pi^\frac{d}{2}}{\left[\prod_{i=1}^d
a_i\right]}\frac{\Gamma\left(s-\frac{d}{2}\right)}{\Gamma(s)}m^{d-2s}\\&+\frac{2\pi^s}{\left[\prod_{i=1}^d
a_i\right]\Gamma(s)} \frac{1}{m^{s-\frac{d}{2}}}\sum_{\mathbf{k}\in
\Z^d\setminus\{ \mathbf{0}\}}\left(\sum_{i=1}^d
\left[\frac{k_i}{a_i}\right]^2\right)^{\frac{2s-d}{4}}
K_{s-\frac{d}{2}} \left(2\pi m \sqrt{\sum_{i=1}^d
\left[\frac{k_i}{a_i}\right]^2}\right),\nonumber
\end{align}where $K_{\nu}(z)$ is the modified Bessel function.
The second term is an analytic function of $s$. The first term shows
that when $d$ is odd, $Z_{d}(s;a_1,\ldots, a_d;m)$ has simple poles
at $s=\frac{d}{2}-j, j=0, 1, 2, \ldots$ with residues
\begin{align}\label{eq6_19_10}
\text{Res}_{s=\frac{d}{2}-j}Z_{d} (s;a_1,\ldots, a_d;
m)=\frac{(-1)^j}{j!}\frac{\pi^{\frac{d}{2}}}{\prod_{i=1}^da_i}\frac{m^{2j}}{\Gamma\left(
\frac{d}{2}-j\right)}.
\end{align}When $d$ is even, $Z_{d}(s;a_1,\ldots, a_d;m)$ only has poles at
$s=\frac{d}{2}-j$ where $j=0, 1, \ldots, \frac{d-2}{2}$, with the
same residue \eqref{eq6_19_10}. By moving the contour of integration
from the line $\text{Re}\; w = c_0, c_0 > d$ to the line
$\text{Re}\; w = -2+\vep$, we obtain for $d=1, 2, 3$,
\begin{align*}
K(\lambda; L_1; m) =& \frac{L_1}{\pi}\lambda^{-1}+\frac{\lambda m^2
L_1}{2\pi}\log \lambda+\frac{\lambda m^2 L_1}{2\pi}\left(\log m
-\log 2-\frac{5}{4}+\gamma\right)\\&+\frac{2\lambda
m}{\pi}\sum_{k=1}^{\infty} k^{-1}K_1(mkL_1)+O(\lambda^2);
\end{align*}
\begin{align*}
K(\lambda; L_1, L_2;m)=&\frac{L_1L_2}{2\pi}\lambda^{-2}-
\frac{L_1L_2}{4\pi}m^2+\frac{L_1L_2}{6\pi}\lambda
m^3+\frac{L_1L_2}{\sqrt{2\pi^3}}\lambda m^{3/2}\\
&\times \sum_{\mathbf{k}\in \Z^2\setminus\{
\mathbf{0}\}}\left(\sum_{i=1}^2
\left[k_iL_i\right]^2\right)^{-\frac{3}{4}} K_{3/2} \left( m
\sqrt{\sum_{i=1}^2 \left[k_iL_i\right]^2}\right)+O(\lambda^2);
\end{align*}and
\begin{align*}
K(\lambda; L_1, L_2,
L_3;m)=&\frac{L_1L_2L_3}{\pi^2}\lambda^{-3}-\frac{L_1L_2L_3}{4\pi^2}m^2\lambda^{-1}-\frac{L_1L_2L_3}{16\pi^2}m^4
\lambda  \log \lambda\\&-\frac{L_1L_2L_3}{16\pi^2} m^4 \lambda\left(
-\frac{3}{2}+\gamma+\log m-\log 2
\right)+\frac{L_1L_2L_3}{2\pi^2}\lambda m^2\\&
\times\sum_{\mathbf{k}\in \Z^3\setminus\{
\mathbf{0}\}}\left(\sum_{i=1}^3 \left[k_iL_i\right]^2\right)^{-1}
K_{2} \left( m \sqrt{\sum_{i=1}^3
\left[k_iL_i\right]^2}\right)+O(\lambda^2),
\end{align*}where $\gamma$ is the Euler constant. Using \eqref{eq6_19_5}, we find that the cut-off
dependent Casimir energy $E_{\text{Cas}}^P\left(\lambda; L_1,
\ldots;  L_3;m\right)$ for $d=1, 2$ and $3$ are given respectively
by
\begin{align*}
E_{\text{Cas}}^P(\lambda; L_1; m) =&
\frac{L_1}{2\pi}\lambda^{-2}-\frac{  m^2
L_1}{4\pi}\log\lambda-\frac{ m^2 L_1}{4\pi}\left(\log m -\log
2-\frac{1}{4}+\gamma\right)\\&-\frac{ m}{\pi}\sum_{k=1}^{\infty}
k^{-1}K_1(mkL_1)+O(\lambda);
\end{align*}
\begin{align*}
E_{\text{Cas}}^P(\lambda; L_1,
L_2;m)=&\frac{L_1L_2}{2\pi}\lambda^{-3}- \frac{L_1L_2}{12\pi}
m^3-\frac{L_1L_2}{\sqrt{8\pi^3}} m^{3/2}\\
&\times\sum_{\mathbf{k}\in \Z^2\setminus\{
\mathbf{0}\}}\left(\sum_{i=1}^2
\left[k_iL_i\right]^2\right)^{-\frac{3}{4}} K_{3/2} \left( m
\sqrt{\sum_{i=1}^2 \left[k_iL_i\right]^2}\right)+O(\lambda);
\end{align*}and
\begin{align*}
E_{\text{Cas}}^P\left(\lambda; L_1, L_2, L_3; m\right)=&\frac{3
L_1L_2L_3}{2\pi^2}\lambda^{-4}-\frac{L_1L_2L_3}{8\pi^2}m^2\lambda^{-2}+\frac{L_1L_2L_3}{32\pi^2}m^4
\log \lambda\\&+\frac{L_1L_2L_3}{32\pi^2} m^4 \left(
-\frac{1}{2}+\gamma+\log m-\log 2 \right)-\frac{L_1L_2L_3}{4\pi^2}
m^2\\& \times\sum_{\mathbf{k}\in \Z^3\setminus\{
\mathbf{0}\}}\left(\sum_{i=1}^3 \left[k_iL_i\right]^2\right)^{-1}
K_{2} \left( m \sqrt{\sum_{i=1}^3
\left[k_iL_i\right]^2}\right)+O(\lambda).
\end{align*} One can show that
in the massless limit ($m=0$), these results coincide with the
corresponding results obtained in \cite{6, 7}.

\section{Casimir effect for massive scalar fields in rectangular
cavities by zeta regularization method, and the attractive/repulsive
nature of the Casimir force} Using zeta regularization method, the
Casimir energy $\hat{E}_{\text{Cas}}^{P}( L_1, \ldots, L_d;m)$
\eqref{eq6_19_2} for a massive scalar field with periodic bc inside
a $d$-dimensional rectangular cavity is defined as \cite{32}
\begin{align*}
\hat{E}_{\text{Cas}}^{P}( L_1, \ldots,
L_d;m):=\frac{1}{4}\lim_{\vep\rightarrow 0}\left(
\zeta_P\left(-\frac{1}{2}+\vep\right)
+\zeta_P\left(-\frac{1}{2}-\vep\right)\right),
\end{align*}where $\zeta_P(s)$ is the  zeta function
\begin{align*}
\zeta_P(s)
=&\mu^{1+2s}\sum_{\mathbf{k}\in\Z^d}(\omega_{\mathbf{k}}^P)^{-2s}=\mu^{1+2s}
Z_d\left(s; \frac{2\pi}{L_1},\ldots,\frac{2\pi}{L_d};m\right);
\end{align*}and $\mu$ is a normalization constant. Using
\eqref{eq6_26_1}, we find that $\hat{E}_{\text{Cas}}^{P}( L_1,
\ldots, L_d;m)$ depends on the normalization constant $\mu$ if and
only is $d$ is odd. For $d=1, 2, 3$, $\hat{E}_{\text{Cas}}^{P}( L_1,
\ldots, L_d;m)$ is given explicitly by
\begin{align*}
\hat{E}_{\text{Cas}}^{P}( L_1; m)=
\frac{L_1}{4\pi}m^2\log\mu-\frac{m^2L_1}{4\pi}\left(\log m-\log
2-\frac{1}{4}\right)-\frac{ m}{\pi}\sum_{k=1}^{\infty}
k^{-1}K_1(mkL_1),
\end{align*}
\begin{align*}
\hat{E}_{\text{Cas}}^P( L_1, L_2;m)=&- \frac{L_1L_2}{12\pi}
m^3-\frac{L_1L_2}{\sqrt{8\pi^3}} m^{3/2}\\
&\times\sum_{\mathbf{k}\in \Z^2\setminus\{
\mathbf{0}\}}\left(\sum_{i=1}^2
\left[k_iL_i\right]^2\right)^{-\frac{3}{4}} K_{3/2} \left( m
\sqrt{\sum_{i=1}^2 \left[k_iL_i\right]^2}\right),
\end{align*}
\begin{align*}
\hat{E}_{\text{Cas}}^P( L_1, L_2, L_3;m)=&-\frac{L_1L_2L_3}{32\pi^2}
m^4 \log\mu+\frac{L_1L_2L_3}{32\pi^2} m^4 \left( \log m-\log 2
-\frac{1}{2}\right)\\&-\frac{L_1L_2L_3}{4\pi^2}
m^2\sum_{\mathbf{k}\in \Z^3\setminus\{
\mathbf{0}\}}\left(\sum_{i=1}^3 \left[k_iL_i\right]^2\right)^{-1}
K_{2} \left( m \sqrt{\sum_{i=1}^3 \left[k_iL_i\right]^2}\right).
\end{align*}
Compare to the result obtained by cut-off regularization method, we
find that the zeta regularized Casimir energy
$\hat{E}_{\text{Cas}}^{P}( L_1, \ldots, L_d;m)$ will agree with the
$\lambda$-independent part of the cut-off dependent Casimir energy
$E_{\text{Cas}}^{P}( \lambda; L_1, \ldots, L_d;m)$ if and only if we
take $\mu =e^{-\gamma}$, where $\gamma$ is the Euler constant. Using
this prescription, we find that  the zeta regularized
Casimir force acting in the $x_1$ direction for $d=3$ is
\begin{align}\label{eq6_27_1}
&\hat{F}_{1,\text{Cas}}^{P}(L_1, L_2, L_3; m) =-\frac{\pa}{\pa
L_1}\hat{E}_{\text{Cas}}^P( L_1, L_2, L_3;m)\\
=&-\frac{L_2L_3}{32\pi^2} m^4 \left( \log m-\log 2
-\frac{1}{2}+\gamma\right)\nonumber\\&+\frac{L_2L_3}{4\pi^2}
m^2\sum_{(k_2,k_3)\in \Z^2\setminus\{
\mathbf{0}\}}\left(\sum_{i=2}^3 \left[k_iL_i\right]^2\right)^{-1}
K_{2} \left( m \sqrt{\sum_{i=2}^3 \left[k_iL_i\right]^2}\right)\nonumber\\
&+\frac{\pa }{\pa L_1} \left(\frac{L_1L_2L_3}{4\pi^2}
m^2\sum_{\substack{k_1\in \Z\setminus\{0\}\nonumber\\
(k_2, k_3)\in \Z^2}}\left(\sum_{i=1}^3
\left[k_iL_i\right]^2\right)^{-1} K_{2} \left( m \sqrt{\sum_{i=1}^3
\left[k_iL_i\right]^2}\right)\right).\nonumber
\end{align}As is shown in Section 3, the last term in \eqref{eq6_27_1} is the net
contribution to the Casimir force acting on the Casimir piston. For
any fixed $m, L_2, L_3$, it is always negative and increase from
$-\infty$ (when $L_1\rightarrow 0^+$) to $0$ (when $L_1\rightarrow
\infty$). On the other hand, the first and second term in
\eqref{eq6_27_1} are independent of $L_1$. Moreover, the second term
is always positive and by taking $m$ sufficiently small, the first
term is also positive. Therefore, for fixed $L_2, L_3$, when $m$ is
sufficiently small and $L_1$ is sufficiently large, the zeta
regularized Casimir force $\hat{F}_{1,\text{Cas}}^P(L_1, L_2, L_3;
m)$ becomes repulsive. Similar reasonings can be used to show that
the zeta regularized Casimir force $\hat{F}_{1,
\text{Cas}}^{D/N}(L_1, L_2, L_3; m)$ acting in the $x_1$ direction
of a rectangular cavity due to a massive scalar field with Dirichlet
and Neumann boundary conditions can be either attractive or
repulsive depending on the values of $L_1, L_2, L_3$ and $m$.

We would also like to remark that in some of the regularization
schemes, the  Casimir force on a wall in a rectangular cavity is
defined by subtracting the Casimir force in the absence of boundary.
In this case, the Casimir force is equal to the sum of the last two
terms in \eqref{eq6_27_1}, i.e. the first term is absent. We call
the
 Casimir force defined in this way the reduced Casimir force. Notice
 that in the massless case, there is no difference between the
 Casimir force and reduced Casimir force in a rectangular cavity. On
 the other hand, it is easy to see that the argument given above shows that the reduced Casimir force in a
 rectangular cavity can be positive or negative, under different
 boundary conditions. In Figure 8, we demonstrate the Casimir force on the
 piston, the Casimir force (and reduced Casimir force) due to the
 interior and exterior regions on a single graph, for the periodic
 bc and Dirichlet bc cases.

\section{An alternative expression for the Casimir force when $a$ is
small} To study the behavior of the Casimir force when $a$ is small,
we first split the summation over $(k_2, \ldots, k_d)\in\Z^{d-1}$ in
\eqref{eq6_20_10} into the term $(k_2,\ldots, k_d)=0$ and the
summation over $(k_2, \ldots, k_d)\in\Z^{d-1}\setminus\{0\}$. Then
applying the identity \eqref{eq6_23_1} to the summation over $k_1\in
\mathbb{Z}\setminus\{0\}$, we find that
\begin{align}\label{eq6_27_6}
&R_d(a,L_2, \ldots,L_d; m)
\\=&-\frac{a\prod_{i=2}^{d}L_i}{2^{d+1}\pi^{\frac{d+1}{2}}}\int_0^{\infty}
t^{-\frac{d+3}{2}}\sum_{k_1=1}^{\infty}\exp\left(-tm^2-\frac{k_1^2a^2}{4t}\right)dt\nonumber\\&+
\frac{a\prod_{i=2}^dL_i}{2^{d+2}\pi^{\frac{d+1}{2}}}\int_0^{\infty}
t^{-\frac{d+3}{2}}\sum_{ (k_2, \ldots, k_d) \in
\Z^{d-1}\setminus\{0\}}\exp\left(
- tm^2 -\frac{1}{4t}\sum_{i=2}^d \left[k_iL_i\right]^2\right)dt\nonumber\\
&- \frac{\prod_{i=2}^dL_i}{2^{d+1}\pi^{\frac{d}{2}}}\int_0^{\infty}
t^{-\frac{d+2}{2}}\sum_{\substack{k_1\in\Z\\(k_2,\ldots,
k_d)\in\Z^{d-1}\setminus\{0\}}}\exp\left( - t\left(m^2+\frac{4\pi^2
k_1^2}{a^2}\right) -\frac{1}{4t}\sum_{i=2}^d
\left[k_iL_i\right]^2\right)dt.\nonumber
\end{align}Taking derivative with respect to $a$ gives
\begin{align}\label{eq6_27_5}
&-\frac{\pa}{\pa a}R_d(a,L_2, \ldots,L_d; m)
\\=&\frac{\prod_{i=2}^{d}L_i}{2^{d+1}\pi^{\frac{d+1}{2}}}\int_0^{\infty}
t^{-\frac{d+3}{2}}\sum_{k_1=1}^{\infty}\exp\left(-tm^2-\frac{k_1^2a^2}{4t}\right)dt\nonumber\\
&-\frac{a^2\prod_{i=2}^{d}L_i}{2^{d+2}\pi^{\frac{d+1}{2}}}\int_0^{\infty}
t^{-\frac{d+5}{2}}\sum_{k_1=1}^{\infty}k_1^2\exp\left(-tm^2-\frac{k_1^2a^2}{4t}\right)dt\nonumber\\&-
\frac{\prod_{i=2}^dL_i}{2^{d+2}\pi^{\frac{d+1}{2}}}\int_0^{\infty}
t^{-\frac{d+3}{2}}\sum_{ (k_2, \ldots, k_d) \in
\Z^{d-1}\setminus\{0\}}\exp\left(
- tm^2 -\frac{1}{4t}\sum_{i=2}^d \left[k_iL_i\right]^2\right)dt\nonumber\\
&+
\frac{\pi^{2-\frac{d}{2}}\prod_{i=2}^dL_i}{2^{d-2}a^3}\int_0^{\infty}
t^{-\frac{d}{2}}\sum_{\substack{k_1\in\Z\\(k_2,\ldots,
k_d)\in\Z^{d-1}\setminus\{0\}}}k_1^2\exp\left( -
t\left(m^2+\frac{4\pi^2 k_1^2}{a^2}\right) -\frac{1}{4t}\sum_{i=2}^d
\left[k_iL_i\right]^2\right)dt.\nonumber
\end{align}The expressions for the Casimir force $F_{\text{Cas}}^{P/D/N}(a; L_1, L_2, L_3;m)$
can then be obtained by using \eqref{eq6_23_5}, \eqref{eq6_23_2} and
\eqref{eq6_23_3}. The results are given by \eqref{eq6_24_1} and
\eqref{eq6_24_2}.

The last  term on the right hand side of \eqref{eq6_27_5} tends to
zero exponentially fast when $a\rightarrow 0^+$. The third term is
independent of $a$. For the first two terms, their asymptotic
behaviors  when $a\rightarrow 0^+$ are not obvious. Let us call the
sum of these two terms as $\mathfrak{T}(a; L_2, \ldots, L_d; m)$.
Here we derive its asymptotic behavior when $a\rightarrow 0^+$. From
the derivation above, we have
\begin{align*}
\mathfrak{T}(a; L_2, \ldots, L_d; m)=&\frac{\pa}{\pa
a}\left\{\frac{a\prod_{i=2}^{d}L_i}{2^{d+1}\pi^{\frac{d+1}{2}}}\int_0^{\infty}
t^{-\frac{d+3}{2}}\sum_{k_1=1}^{\infty}\exp\left(-tm^2-\frac{k_1^2a^2}{4t}\right)dt\right\}\\
=&\frac{\prod_{i=2}^{d}L_i}{2^{d+1}\pi^{\frac{d+1}{2}}}\frac{\pa}{\pa
a}\mathfrak{P}\left(-\frac{d+1}{2}; a;m\right),
\end{align*}where $\mathfrak{P}(s; a; m)$ is the analytic function
(of $s$) defined by
\begin{align*}
\mathfrak{P}(s; a; m)=a\int_0^{\infty}
t^{s-1}\sum_{k=1}^{\infty}\exp\left(-tm^2-\frac{k^2a^2}{4t}\right)dt.
\end{align*}Making a change of variable $t\mapsto  a^2t$, we have
\begin{align*}
\mathfrak{P}(s; a; m)=a^{1+2s}\int_0^{\infty}
t^{s-1}\sum_{k=1}^{\infty}\exp\left(-ta^2m^2-\frac{k^2}{4t}\right)dt.
\end{align*}For $s>1/2$,  Jacobi inversion formula
\eqref{eq6_23_1} gives
\begin{align*}
\mathfrak{P}(s; a; m)=&a^{1+2s}\int_0^{\infty}
t^{s-1}\sum_{k_1=1}^{\infty}e^{-ta^2m^2}\left\{-\frac{1}{2} +
\sqrt{\pi t}\sum_{k=-\infty}^{\infty} e^{ -4\pi^2k^2t}
\right\}dt.\end{align*} When $a$ is small, we can use the Taylor
expansion of $e^{-ta^2m^2}$ to get\begin{align*}
\mathfrak{P}(s; a; m)=&-\frac{am^{-2s}}{2}\Gamma(s)+\sqrt{\pi}m^{-1-2s}\Gamma\left(s+\frac{1}{2}\right)\\
&+2\sqrt{\pi}a^{1+2s}
\sum_{j=0}^{\infty}\frac{(-1)^j}{j!}a^{2j}m^{2j}\int_{0}^{\infty}
t^{s+j-1/2} \sum_{k=1}^{\infty}e^{ -4\pi^2k^2t}dt\\
=&-\frac{am^{-2s}}{2}\Gamma(s)+\sqrt{\pi}m^{-1-2s}\Gamma\left(s+\frac{1}{2}\right)\\
&+2\sqrt{\pi}a^{1+2s}
\sum_{j=0}^{\infty}\frac{(-1)^j}{j!}\frac{a^{2j}m^{2j}}{(2\pi)^{2j+2s+1}}\Gamma\left(s+j+\frac{1}{2}
\right)\zeta(2s+2j+1),
\end{align*}where $\zeta(s)$ is the Riemann zeta function. One can check that the formula on the right hand side of
the last equality define an analytic function of $s$. Putting
$s=-(d+1)/2$ and taking derivative with respect to $a$, we find that
if $d$ is odd,
\begin{align*}
&\mathfrak{T}(a; L_2, \ldots, L_d; m)\\
=& 2\pi^{d/2}a^{-d-1}
\left[\prod_{i=2}^d L_i\right] \sum_{\substack{j\in \mathbb{N}\cup\{0\}\\
j\neq \frac{d+1}{2}}}\frac{(-1)^j}{j!} (2\pi)^{-2j}m^{2j}a^{2j}
\Gamma\left(j+1-\frac{d}{2}\right)\zeta(2j-d)\\
&+\frac{(-1)^{\frac{d+1}{2}}}{\left(\frac{d+1}{2}\right)!}\frac{m^{d+1}}{2^{d+2}\pi^{\frac{d+1}{2}}}
\left[\prod_{i=2}^dL_i\right]\left\{
2\log\frac{am}{4\pi}+2-\psi\left(\frac{d+3}{2}\right)+\gamma\right\};
\end{align*}and when $d$ is even,
\begin{align*}
&\mathfrak{T}(a; L_2, \ldots, L_d; m)\\
=&2\pi^{d/2}a^{-d-1}
\left[\prod_{i=2}^d L_i\right] \sum_{\substack{j\in \mathbb{N}\cup\{0\}\\
j\neq \frac{d}{2}}}\frac{(-1)^j}{j!} (2\pi)^{-2j}m^{2j}a^{2j}
\Gamma\left(j+1-\frac{d}{2}\right)\zeta(2j-d)\\
&-\frac{\Gamma\left(-\frac{d+1}{2}\right)}{2^{d+2}\pi^{\frac{d+1}{2}}}m^{d+1}\left[\prod_{i=2}^d
L_i\right] -\frac{(-1)^{\frac{d}{2}}}{\left(
\frac{d}{2}\right)!}\frac{m^d}{2^d\pi^{\frac{d}{2}}}\left[\prod_{i=2}^d
L_i\right]a^{-1}.\end{align*}Here $\psi(z)$ is the logarithm
derivative of the gamma function, and for $s\leq 0$,
$\Gamma(s/2+1)\zeta(s)$ is understood as
\begin{align*}
\Gamma\left(\frac{s+2}{2}\right)\zeta(s)
=\frac{\pi^{s-\frac{1}{2}}}{2}s\Gamma\left(\frac{1-s}{2}\right)\zeta(1-s).
\end{align*}When $d=1, 2, 3$, using $\zeta(2)=\pi^2/6$ and $\zeta(4)=\pi^4/90$,  we have respectively
\begin{align*}
&\mathfrak{T}(a; m)=-\frac{\pi}{6a^2}-\frac{\pi m^2}{4}\log a +
O(a^0);
\end{align*}
\begin{align*}
&\mathfrak{T}(a; L_2;m)=-\frac{L_2}{\pi a^3}\zeta(3)
+\frac{m^2L_2}{4\pi a}+O(a^0);
\end{align*}and
\begin{align*}
&\mathfrak{T}(a; L_2, L_3;m)=-\frac{\pi^2L_2L_3}{30 a^4}
+\frac{m^2L_2L_3}{24a^2}+\frac{\pi^2m^4}{32}L_2L_3\log a+O(a^0).
\end{align*}

\end{document}